%% file: arXiv.PLB_34jet.tex
\documentclass[preprint,3p,twocolumn,number,sort&compress]{elsarticle}

\usepackage{pslatex}
\usepackage{psfrag}
\usepackage{slashed}
\usepackage{xspace}
\usepackage{color}
\usepackage{amsmath}
\usepackage{amssymb}

\usepackage{hyperref}
\hypersetup{bookmarksnumbered,}

\graphicspath{{figures/}{jetdistr/}}

\newcommand{\ngluon}{{\sc NGluon}\xspace}
\newcommand{\NJet}{{\sc NJet}\xspace}
\newcommand{\BlackHat}{{\sc BlackHat}\xspace}

\definecolor{mygreen}{rgb}{0,0.7,0}

\newcommand{\qbar}{{\overline{q}}}

\def\as{\ensuremath{\alpha_s}}
\def\mur{\ensuremath{\mu_r}}
\def\muf{\ensuremath{\mu_f}}
\def\LO{\mbox{\scriptsize LO}}
\def\NLO{\mbox{\scriptsize NLO}}
\def\njet{n\text{-jet}}

\def\Eq#1{Eq.~(\ref{#1})}

\def\Fig#1{Fig.~\ref{#1}}
\def\Tab#1{Tab.~\ref{#1}}
\def\Ref#1{Ref.~\cite{#1}}
\def\Refs#1{Refs.~\cite{#1}}

\title{%
  \begin{picture}(0,0)
    \put(420,100){{\normalsize HU-EP-12/28} }
  \end{picture}%
  NLO QCD corrections to multi-jet production at the LHC with a
  centre-of-mass energy of $\sqrt{s}=8$ TeV}

\author[a]{Simon Badger}
\ead{badger@nbi.dk}
\author[b]{Benedikt Biedermann}
\ead{benedikt.biedermann@physik.hu-berlin.de}
\author[b]{Peter Uwer}
\ead{peter.uwer@physik.hu-berlin.de}
\author[a]{Valery Yundin}
\ead{yundin@nbi.dk}

\address[a]{
Niels Bohr International Academy and Discovery Center,
The Niels Bohr Institute,\\%
University of Copenhagen, Blegdamsvej 17, DK-2100 Copenhagen, Denmark}
\address[b]{
Humboldt-Universit\"at zu Berlin, Institut f\"ur Physik,\\%
Newtonstra{\ss}e 15, D-12489 Berlin, Germany}

\begin{document}

\begin{abstract}
  We study three and four jet production in hadronic collisions
  at next-to-leading order accuracy in massless QCD. We cross check
  results previously obtained by the \BlackHat collaboration for the LHC
  with a centre-of-mass energy of $\sqrt{s}=7$ TeV and present new
  results for the LHC operating at 8 TeV. We find large negative NLO
  corrections reducing the leading-order cross sections by about
  40--50\%. Furthermore we observe an important reduction of the
  scale uncertainty. In addition to the cross sections we also
  present results for differential distributions. The dynamical
  renormalization/factorization scale used in the calculation leads to
  a remarkably stable K-factor.
  The results
  presented here were obtained with the \NJet package \cite{Badger:2012tobedone}, a publicly
  available library  for the evaluation of one-loop amplitudes in
  massless QCD.
  \vspace{\baselineskip}\\
  Draft from \today
\end{abstract}


\maketitle

\section{Introduction}
Multi-jet production in hadronic collisions via the strong interaction
represents an important testing ground for quantum chromodynamics (QCD).
Jet production in massless QCD may provide valuable information to
constrain the QCD coupling constant $\as$ and/or the parton
distribution functions (PDFs).  Furthermore owing to their significant
production rates multi-jet processes may contribute as background
processes to a large variety of interesting signal reactions including
processes relevant for new physics searches.  Independent of whether
multi-jet production in QCD is studied as signal or background process
precise theoretical predictions are mandatory.

In the past substantial progress has been made in the numerical
evaluation of leading order (LO) matrix elements for large
multiplicities.  Using publicly available tools like
Alpgen~\cite{Mangano:2002ea}, Sherpa+Comix~%
\cite{Gleisberg:2008ta,Gleisberg:2008fv}, MadGraph/MadEvent~%
\cite{Alwall:2007st,Maltoni:2002qb,Alwall:2011uj} and Helac \cite{Cafarella:2007pc} multiplicities with
up to~12 final state jets can be calculated. However leading-order
cross section predictions suffer from a substantial (residual) dependence
on the unphysical renormalization scale $\mur$. This is in particular
true for high multiplicities due to the high power in \as\ occurring
in the perturbative expansion. For a process starting with $\as^n$
the scale dependent terms are proportional to
\begin{equation}
  n \beta_0 \as \sigma_0
\end{equation}
where $\sigma_0$ denotes the leading-order cross section and $\beta_0$
is the leading coefficient of the QCD beta function. Precise
theoretical predictions thus require at least next-to-leading order
(NLO) corrections in the coupling \as\ to reduce the scale dependence.
While the NLO corrections to two jet production were calculated
already in 1992 \cite{Ellis:1992en,Giele:1993dj}  and the
corrections to three jet production a decade later
\cite{Nagy:2001fj,Nagy:2003tz} (results for the gluon channel were
obtained already in \Refs{Kilgore:1996sq,Trocsanyi:1996by})
further progress has been hindered by both the increasing complexity
of the one-loop amplitudes and the large number of tree-level
processes entering the infra-red counter-terms in the Catani-Seymour
subtraction formalism \cite{Catani:1996jh}. The latter problem could
be solved by an automated generation of subtraction terms based on the
aforementioned tree-level technology
\cite{Gleisberg:2007md,Seymour:2008mu,Frederix:2008hu,Hasegawa:2009tx,Czakon:2009ss,Frederix:2009yq}.  While
technically solved the available computer resources still impose a
severe restriction on the multiplicity which can be handled in
practice. As far as the virtual corrections are concerned just the
mere number of Feynman diagrams in the conventional Feynman diagram
based approach e.g. $\sim 15,000$ for $gg\to 4g$ at one-loop
(neglecting self-energy type corrections) shows that
alternative techniques are required.  In recent years new methods for
the evaluation of one-loop amplitudes have relieved the long standing
bottleneck in producing virtual corrections for processes with high
parton multiplicities. The on-shell unitarity method
\cite{Bern:1994zx,Bern:1994cg} and the subsequently developed
generalized unitarity method
\cite{Britto:2004nc,Forde:2007mi,Ellis:2007br,Giele:2008ve,
  Badger:2008cm,Berger:2008sj} offer an interesting alternative to
conventional Feynman diagram based methods. In particular the
unitarity method allows to use leading order amplitudes as basic
building blocks for the one-loop calculation. It is thus possible to
make use of the efficient tree-level machinery mentioned above.  In
parallel with the development of the unitarity method also the
conventional approach has been continuously improved in the last couple of
years and a mixture of methods are currently exploited
\cite{Cascioli:2011va}. 
We also mention that recently other directions have been
investigated which offer the potential to become powerful alternatives
to those above \cite{Becker:2011vg}.
While improvements are still ongoing, NLO technology
is already now automated and flexible enough
to cover a wide range of processes for four final state particles
\cite{Berger:2009zg,Berger:2009ep,Berger:2010vm, Bredenstein:2009aj,
  Bredenstein:2010rs,Bevilacqua:2009zn,
  Bevilacqua:2010ve,Denner:2010jp,Bevilacqua:2010qb,Bevilacqua:2011aa,
  Denner:2012yc,Bevilacqua:2012em,
  Binoth:2009rv,Greiner:2011mp,Greiner:2012im,Melnikov:2009wh,
  KeithEllis:2009bu,Melia:2011dw,Melia:2010bm,
  Campanario:2011cs,Campanario:2011ud} and a handful of $2\to
\geq 5$ processes
\cite{Berger:2010zx,Ita:2011wn,Frederix:2010ne,Becker:2011vg}. A
number of automated approaches to the computation of virtual
amplitudes have appeared as public codes
\cite{Badger:2010nx,Hirschi:2011pa,Bevilacqua:2011xh,Cullen:2011ac}.

Despite the aforementioned progress only very recently first results
for four jet production in NLO accuracy became available.  In Ref.
\cite{Bern:2011ep} the \BlackHat collaboration published results for
three and four jet production in NLO accuracy for the LHC running at 7~TeV
centre-of-mass energy. Given the complexity of the calculation it is
very important to perform an independent cross check. Furthermore we
present results for the LHC running at 8~TeV and discuss various
differential distributions.  For the evaluation of the virtual matrix
elements we use the publicly available package \ngluon
\cite{Badger:2010nx}.  \ngluon\ uses on-shell methods to evaluate
numerically one-loop primitive amplitudes in pure gauge theory.
Recently we have extended \ngluon\ to account for amplitudes involving
massless quarks \cite{Badger:2011zv,Badger:2012dd,
  Badger:2012tobedone}. We note that in \Refs{Badger:2011zv,Badger:2012dd,
  Badger:2012tobedone} no approximation in the colour is performed.
The real corrections and the cancellation of
soft- and collinear divergences are obtained within the Catani-Seymour
dipole subtraction method \cite{Catani:1996jh}.  For this contribution
we use the Sherpa Monte-Carlo event generator \cite{Gleisberg:2007md}.
We will briefly comment on some technical aspects in the next section.
In section~3 we will present the comparison with the results published
by \BlackHat \cite{Bern:2011ep}.  In addition we show new results for
8~TeV for three and four jet production.  Our conclusions are
presented in section~4.

\section{Outline of the calculation}
In what follows we briefly outline the calculation. We work in
massless QCD with 5 flavours. In particular we include the bottom quark in the
initial state. The top-quark is assumed to be
integrated out. Through the matching of the coupling constants between
the five and the six flavour theories some of the corrections due to
top-quarks are retained. We expect the neglected corrections to be small.
For three jet production the contributing processes are given by all
possible crossings of the following transitions:
\begin{equation}
  0 \to g g g g g,\quad
  0 \to q \qbar g g g,\quad
  0 \to q \qbar q' \qbar' g,
  \label{eq:3jsubprocesses}
\end{equation}
where $q$ and  $q'$ denote a generic quark. For the four jet
production the corresponding processes are derived from
\begin{equation}
  0\to g g  g g g g,
\end{equation}
and
\begin{equation}
  0\to q \qbar g g g g,\quad
  0\to q \qbar q' \qbar'  g g,\quad
  0\to q \qbar q' \qbar' q'' \qbar''.
  \label{eq:4jsubprocesses}
\end{equation}
In principle we need to distinguish the like-flavour cases from the
case that all quark flavours are different. However technically the
like-flavour processes can be obtained from the latter by an
appropriate (anti) symmetrization with respect to the quark momenta.
For example the amplitude $A_{q\qbar q\qbar g}$ for the transition
$0\to q \qbar q \qbar g$ is obtained through
\begin{eqnarray}
  A_{q\qbar q\qbar g}(1,2,3,4,5) &=&\nonumber \\
  &&\hspace{-3cm}A_{q\qbar q'\qbar' g}(1,2,3,4,5)
  -A_{q\qbar q'\qbar' g}(1,4,3,2,5),
  \label{eq:likefermion}
\end{eqnarray}
where we used the short hand notation to abbreviate with $i$ the
momentum $k_i$ and helicity $\lambda_i$ of the $i$-th parton ($i = (k_i,\lambda_i)$).
The expansion of the $n$-jet differential cross section in the
coupling $\as$ reads:
\begin{equation}
  d\sigma_n = d\sigma_n^{\LO} +  d\delta\sigma_n^{\NLO}
  + {\cal O}(\as^{n+2})
\end{equation}
($d\sigma_n^{\LO}\sim \as^n,  d\delta\sigma_n^{\NLO} \sim \as^{n+1}$)
with the leading order cross section given by
\begin{align}
  d\sigma_n^{\LO} =& \sum_{\substack{i,j\\ \in \{q,\qbar,g\}}} dx_i dx_j
  F_{j/H_2}(x_j,\muf)F_{i/H_1}(x_i,\muf)
  \nonumber \\
  &
  {}\times
  d\sigma_n^{\rm B}\bigl(i(x_iP_1)+j(x_jP_2) \to n \mbox{ part.}\bigr).
  \label{eq:HadXSection}
\end{align}
$P_1, P_2$ denote the momenta of the incoming hadrons.  $x_i$ are the
momentum fractions carried by the initial state partons $i$ with
respect to the incoming hadrons $H_{1}, H_{2}$. The momentum sum of
the incoming partons is thus given by $P=x_i P_1 + x_j P_2$. As usual
we neglect the masses of the incoming hadrons. $F_{i/H}(x_i,\muf)$ are
the parton distribution functions which roughly speaking describe the
probability to find a parton $i$ inside the hadron $H$ with a momentum
fraction between $x_i$ and $x_i + dx_i$. The factorization scale is
denoted by \muf.  The partonic cross section for the reaction $(ij \to
n\mbox{-jets})$ in Born approximation is given by
$d\sigma_n^B\bigl(i(x_1P_1)+j(x_jP_2) \to n\mbox{-jets}\bigr)$.  In terms of the
leading order squared matrix elements $|{\cal M}_n|^2$ the explicit
expression for $d\sigma_n^B$ reads:
\begin{multline}
  d\sigma_n^{\rm B} =
  \frac{1}{2 \hat s}\prod_{\ell =1}^n\frac{d^3k_\ell }{ (2\pi)^3 2E_\ell}
  \Theta_{\njet}
  \\
  {}\times(2\pi)^4
  \delta\Bigl(P-\sum_{m=1}^n k_m\Bigr)\big|{\cal M}_n(ij\to n \mbox{ part.})\big|^2,
  \label{eq:PartonXSection}
\end{multline}
where $k_i$ are the four momenta of the outgoing partons and $\hat s =
2 x_i x_j (P_1\cdot P_2) = (\sum_{i=1}^n k_i)^2$ is the partonic
centre of mass energy squared. The jet algorithm is encoded through
$\Theta_{\njet}$. $\Theta_{\njet}$ is a function of the final state
parton momenta $k_i$. It is one if the corresponding $n$ parton
momenta represent an $n$-jet event and zero otherwise. As mentioned in
the introduction different publicly available tools exist to evaluate
the matrix elements ${\cal M}(ij\to n \mbox{ part.})$. We used
Amegic++ \cite{Krauss:2001iv} within the Sherpa framework to achieve
this task. Cross checks on the matrix elements were performed using
Comix \cite{Gleisberg:2008fv}. To perform the numerical integration
the Sherpa Monte Carlo event generator \cite{Gleisberg:2007md} is
used.

At NLO accuracy we need to consider the one-loop corrections
$d\sigma_n^{\rm V}$ and the real corrections $d\sigma_{n+1}^{\rm R}$
due to an additional parton in the final state. Both contributions
$d\sigma_n^{\rm V}$ and $d\sigma_{n+1}^{\rm R}$ individually contain
collinear and soft singularities. To obtain a finite result the two
contributions must be combined and the initial state singularities
must be factorized. A convenient method to perform the cancellation of
the soft and collinear divergences is the Catani-Seymour subtraction
method \cite{Catani:1996jh}. The basic idea is to introduce local
counter-terms which render the integration of the real corrections
finite and are easy enough so that an analytic integration over the
soft and collinear regions of phase space is possible. Schematically
the total cross section reads:
\begin{multline}
  \delta\sigma^{\NLO} = \int\limits_n
  \big(d\sigma_n^{\rm V}
  + \int\limits_1 d\sigma_{n+1}^{\rm S}\big)
  + \int\limits_n d\sigma_n^{\rm Fac.}
    \\
  + \int\limits_{n+1} \big(d\sigma_{n+1}^{\rm R} - d\sigma_{n+1}^{\rm S}\big),
\end{multline}
where $d\sigma_{n+1}^{\rm S}$ denotes the local counter-term and $d\sigma_n^{\rm Fac.}$ is due to the
factorization of initial state singularities.
It is thus convenient to split the NLO corrections into three contributions:
\begin{align}
  d\delta\sigma_n^{\NLO} &= d\bar\sigma_n^{\rm V} + d\bar\sigma_n^{\rm
    I} + d\sigma_{n+1}^{\rm RS},
  \label{eq:NLOxs}
\end{align}
the finite part of the virtual corrections $d\bar\sigma_n^{\rm V}$,
the finite part of the integrated subtraction terms together with the
contribution from the factorization $d\bar\sigma_n^{\rm I}$ and the
real corrections combined with the subtraction terms
$d\sigma_{n+1}^{\rm RS}$. As in the case of the LO cross sections we
use Sherpa in combination with Amegic++ to calculate
$d\bar\sigma_n^{\rm I}$ and $d\sigma_n^{\rm RS}$. Again cross checks
on the tree-level amplitudes were performed using Comix
\cite{Gleisberg:2008fv}.

The one-loop matrix elements required for the virtual corrections
$d\bar\sigma_n^{\rm V}$ are evaluated using an on-shell generalized
unitarity set-up (see Ref.~\cite{Ellis:2011cr} for a recent review)
for multi-parton primitive amplitudes
\cite{Badger:2010nx,Badger:2011zv,Badger:2012dd}.  In
\Ref{Badger:2010nx} only colour-ordered pure gluon amplitudes were
considered. To account for massless quarks appearing in the loop or as
external partons we have extended the \ngluon package to allow the
computation of primitive amplitudes involving quarks. The primitive
amplitudes correspond to the colour-ordered amplitudes in the pure
gluon case.   Details on the extension of the
\ngluon\ package are given in Refs.
\cite{Badger:2011zv,Badger:2012dd,Badger:2012tobedone}. The entire
library together with additional code to perform the colour algebra
--- on which we comment below --- is publicly available as
\NJet\footnote{To download \NJet visit the project home page at\\
\url{ https://bitbucket.org/njet/njet/}.} package. A detailed
description how to use the library can be found in
\Ref{Badger:2012tobedone}.

For a given process the primitive one-loop amplitudes provide all
required information to reconstruct the full amplitude including the
complete colour information. In particular the partial amplitudes
appearing in the colour decomposition of the full amplitude can be
obtained as linear combinations of primitive amplitudes. Since in our
calculation no approximation in the sum over colour is performed, we
need to express all partial amplitudes in terms of the primitive
amplitudes.  This is in general a non-trivial task. In Ref.~%
\cite{Ellis:2011cr} an algorithm to establish the relation between the
partial amplitudes and the primitive amplitudes has been presented.
The method uses the colour decomposition of the full amplitude in
terms of colour stripped Feynman diagrams. This decomposition is
obtained by separating for each Feynman diagram the Lorentz structure
from the colour structure. On the other hand the primitive
amplitudes can also be expressed as linear combinations of colour
stripped Feynman diagrams. From the comparison of the two
representations the relation between the partial and the primitive
amplitudes can be extracted.  For additional details on the method we
refer to Ref.~\cite{Ellis:2011cr}.  We have applied this method to
produce results for up to seven point amplitudes \cite{Badger:2012dd}.
The explicit formulae have also been implemented in the \NJet\
library.  We note that the relations between the primitive
amplitudes and the partial amplitudes for up to seven partons are also
given in Ref.~\cite{Ita:2011ar}. Since we apply additional
  symmetries we slightly differ in the number of independent primitive
  amplitudes required in the numerical evaluation. To check the
correctness of our implementation of the virtual corrections we have
compared our results for individual phase points as far as possible
with {\sc GoSam} \cite{Cullen:2011ac} and {\sc Helac-NLO}
\cite{Bevilacqua:2011xh}. Furthermore we also checked the benchmark
points provided by \BlackHat \cite{Bern:2011ep}. We obtained at
least an agreement of eight digits. This is largely sufficient for
all practical applications.

To perform the phase space integration we interfaced the \NJet\
library to the Sherpa event generator using the Binoth Les Houches
Accord \cite{Binoth:2010xt}. As a technical remark we add that we used
the {\sc HepMC} file format \cite{Dobbs:2001ck} for weighted events.

\section{Cross sections for multi-Jet production at the LHC}

\subsection{Numerical setup and checks}
Before discussing the results for three and four jet production at the
LHC in massless QCD we briefly describe the numerical setup adopted in
the calculation. As shown in \Eq{eq:HadXSection} we need to specify
the parton distribution functions. We use the MSTW2008 PDF set
\cite{Martin:2009iq}. The MSTW2008 PDF set also provides a set of
error PDFs which can be used to assess the uncertainties due to our
incomplete knowledge of the PDFs.  We have not performed any detailed
analysis of PDF uncertainties or comparison between alternative fits
\Refs{Lai:2010vv,Ball:2012cx,Alekhin:2012ig}, which we leave for a
future publication.  In particular it would be interesting to see how
the MSTW2008 PDF set compares with the ABM12 set
\cite{Alekhin:2012ig}, which comes with a slightly different value for
\as\ and differs significantly in the gluon distribution at large $x$.

For consistency we use the \as\ values as provided by the PDF set. We
note that the MSTW2008 PDF set contains so-called leading-order,
next-to-leading order and next-to-next-to-leading order PDFs. They
were obtained by using different orders in the evolution and different
orders for the hard scattering coefficients in the fits to data. Since
\as\ is extracted together with the PDFs different orders come in
general with
different \as\ values. In particular for the LO set we have $\as^{\rm
  LO}(\mur=m_Z) = 0.13939$ while for the NLO set the corresponding
value is $\as^{\rm NLO}(\mur=m_Z) = 0.12018$.  The large value of
$\as^{\rm LO}$ reflects the fact that LO PDF fits in general don't
work very well since sizeable NLO corrections are not taken into
account. This in turn leads to the large value for \as\ which
partially compensates the missing corrections in the hard scattering
coefficients. Despite this obvious tension with the world average for
$\as$ we follow in our default setup the standard procedure to use LO
PDFs together with $\as^{\rm LO}$ to evaluate LO cross sections. For
the NLO cross sections we use the NLO PDFs together with $\as^{\rm
  NLO}$ everywhere. When we discuss the size of the NLO corrections we
will come back to this point.  The PDFs and \as\ dependent on the
unphysical scales $\muf$ and $\mur$. For the results presented here we
set $\muf=\mur\equiv\mu$.  For distributions where the typical energy
scale $Q$ changes significantly like for example the transverse
momentum distribution of the highest energetic jet a fixed value of
$\mu$ may lead to a poor behaviour of the perturbative expansion due
to the appearance of possible large logarithms of the form
$\ln({\mu / Q})$. In such cases using a phase space dependent $\as$
is usually a better choice since some of the logarithms may be
resummed through the evolution of $\as$. As a consequence we use a
dynamical scale $\mu$ based on the sum of the total transverse
momentum of the final state partons
\begin{gather}
  \hat{H}_T = \sum_{i=1}^{N_\text{parton}} p_{T,i}^\text{parton}.
\end{gather}
In particular we set $\mu=\hat{H}_T/2$.
To estimate the effect of uncalculated higher orders we consider
the scale variation $\hat{H}_T/4\leq\mu\leq \hat{H}_T$.

For the jet algorithm appearing in \Eq{eq:PartonXSection} through
$\Theta_{\njet}$ we use the anti-kt algorithm \cite{Cacciari:2008gp} as
implemented in {\sc fastjet} \cite{Cacciari:2011ma}. The jet-radius
parameter $R$ is set to
\begin{equation}
  R=0.4
\end{equation}
following the value adopted by the ATLAS collaboration.
Events were generated using identical cuts to that of the multi-jet
measurements from ATLAS \cite{Aad:2011tqa} and the recent study by the
\BlackHat collaboration \cite{Bern:2011ep}. In particular the transverse
momentum, $p_T$, of the first jet is required to be larger than $80$ GeV with
subsequent jets required to have at least $p_T>60$ GeV.
Rapidity cuts of $|\eta|<2.8$ were also taken.

To check our setup we reproduced the results quoted in
\Ref{Bern:2011ep}. For the three and four jet cross sections at a
centre-of-mass energy of 7 TeV we find:
\begin{align}
  \sigma_3^{\text{7TeV-LO}} &= 93.40(0.03)^{+50.37}_{-30.34}\: {\rm nb},\\
  \sigma_3^{\text{7TeV-NLO}} &= 53.74(0.16)^{+2.06}_{-20.72}\: {\rm nb},
  \label{eq:3jXS7TeV}
\end{align}
and
\begin{align}
  \sigma_4^{\text{7TeV-LO}} &= 9.98(0.01)^{+7.40}_{-3.93}\: {\rm nb},\\
  \sigma_4^{\text{7TeV-NLO}} &= 5.61(0.13)^{+0.0}_{-2.23}\: {\rm nb},
  \label{eq:4jXS7TeV}
\end{align}
where the NLO cross section $\sigma_n^{\NLO}$ is defined as
\begin{equation}
  \sigma_n^{\NLO} = \sigma_n^{\LO} + \delta \sigma_n^{\NLO}.
\end{equation}
The number in brackets show the Monte-Carlo errors of the numerical
integration and the sub- and super-scripts refer to the
minimum and maximum values obtained through scale variations
estimated at $\mu=\hat{H}_T/4$ and $\mu=\hat{H}_T$.
Comparing the results given in \Eq{eq:3jXS7TeV} and \Eq{eq:4jXS7TeV} we find
perfect agreement with \Ref{Bern:2011ep}. In addition for the $p_T$ distribution of the fourth
leading jet in $pp\to 4$jets we compared to the table presented by the \BlackHat collaboration,
again finding full agreement within Monte-Carlo errors. The results are shown in Table
\ref{tab:4jpt47tev}. We believe that together with
the check of the matrix elements for individual phase space points the
successful comparison with the \Ref{Bern:2011ep} is highly
non-trivial giving us confidence that the results presented in
this paper are correct.

\begin{table}
  \centering
  \begin{tabular}[h]{ccc}
    \hline \\[-3mm]
    $p_T$ (GeV) &  $d\sigma_4^{\text{LO}} / dp_{T,4}$ & $d\sigma_4^{\text{NLO}} / dp_{T,4}$
    \\[1mm]
    \hline \\[-3mm]
    $60$ --- $80$ & $398.6(0.4)^{+295.9}_{-157.0}$ & $223(6)^{+0.0}_{-92}$ \\[1mm]
    $80$ --- $110$ & $57.53(0.07)^{+42.54}_{-22.66}$ & $32.5(1.1)^{+0.0}_{-11.9}$ \\[1mm]
    $110$ --- $160$ & $5.24(0.01)^{+3.87}_{-2.06}$ & $3.1(0.2)^{+0.0}_{-0.8}$ \\[1mm]
    $160$ --- $210$ & $0.394(0.002)^{+0.285}_{-0.156}$ & $0.26(0.02)^{+0.0}_{-0.08}$ \\[1mm]
    \hline
  \end{tabular}
  \caption{A table of values for the differential $p_T$ distribution of the $4^{\text{th}}$ leading jet in
  $pp\to4$ jets at $\sqrt{s}=7$ TeV given in units of pb/GeV. The numbers can be compared directly to those of
  Ref.~\cite{Bern:2011ep}.}
  \label{tab:4jpt47tev}
\end{table}

\subsection{Three Jet Production}
\begin{figure}[t]
  \begin{center}
    \includegraphics[width=\columnwidth]{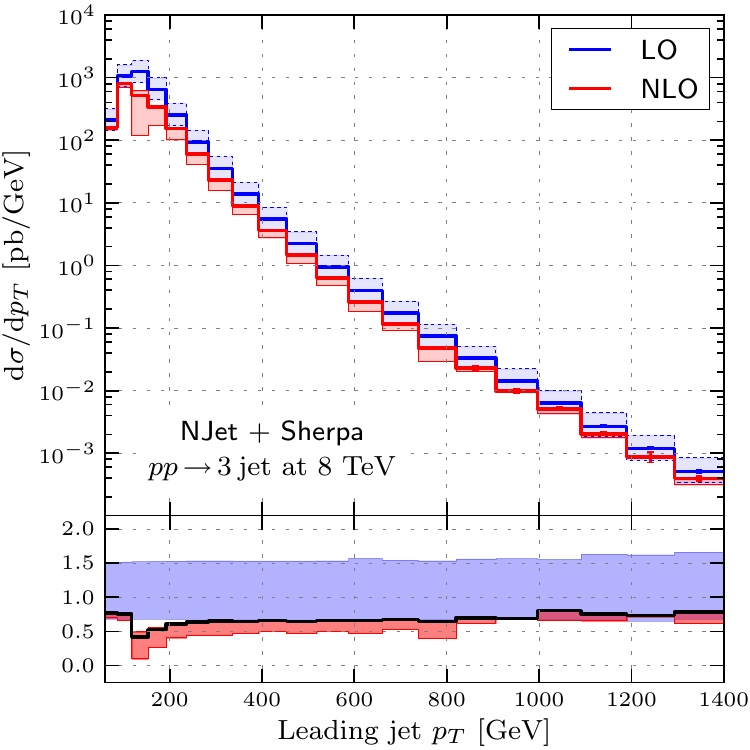}
  \end{center}
  \caption{$p_T$ distribution for the leading jet at the LHC with a
    centre-of-mass energy of 8 TeV.  The upper plots show leading
    order (blue) and next-to-leading order (red) results
    for the central scale $\mu=\hat{H}_T/2$.  The bands show the LO and NLO
    scale variations respectively. In the lower plot we show the ratio
    of LO and NLO together with the LO scale variations (blue band)
    and NLO scale variations (red band).}
  \label{fig:3jpT1}
\end{figure}
We now present results for three jet production at 8 TeV
centre-of-mass energy. Since three jet production has been studied in
some detail before \cite{Nagy:2001fj,Nagy:2003tz} we limit our discussion to a few basic
quantities. Using the aforementioned setup we find for the
total three jet cross section:
\begin{align}
  \sigma_3^{\text{8TeV-LO}} &= 126.65(0.05)^{+66.56}_{-40.40}\: {\rm nb},\\
  \sigma_3^{\text{8TeV-NLO}} &= 72.57(0.16)^{+2.71}_{-28.08}\: {\rm nb},
  \label{eq:3jXS}
\end{align}
again the numbers in parentheses quote the numerical uncertainty due
to the Monte Carlo integration while the sub- and super-scripts show
the effect of the scale variation. We note a significant reduction of
the scale uncertainty similar to what has been observed for 7 TeV.
Furthermore the NLO corrections give a sizeable change of the cross
section prediction: The NLO results are reduced by about 40\% with
respect to the LO cross section again in perfect agreement to what has
been observed for 7 TeV in \Ref{Bern:2011ep}. Compared to a collider
energy of 7 TeV the cross sections are about 35\% larger as a
consequence of the increased parton fluxes at 8 TeV.  As a technical
side remark we point out that the numerical integration is very well
under control: The numerical uncertainty for the central value amounts
to about 2 per mille.
In \Fig{fig:3jpT1}
we show results for the $p_T$ distribution of the leading jet.
As for the total three jet corrections we observe a sizeable reduction
of the NLO prediction with respect to the leading order result. The
scale uncertainty is reduced to about 25\% of the LO scale
uncertainty. With the exception of the low $p_T$ region we observe a rather
constant K-factor. The dynamical scale chosen for the renormalization
and factorization scale indeed avoids large logarithmic corrections at
large $p_T$. For small $p_T$ we expect that soft gluon corrections
become important. A reliable prediction in that region would thus
require to go beyond a fixed order calculation.\footnote{Strictly
  speaking already the dynamical scale setting procedure is beyond a
  fixed order calculation. A certain class of possible large
  logarithms are resummed via the renormalization group. In case of
  soft logarithms in addition soft gluon resummation would be required.}
We have also studied the $p_T$ distribution for the
second and the third jet in the $p_T$ ordering. The results are very
similar to what is shown in \Fig{fig:3jpT1}. Again we find a rather
flat K-factor. The main difference to the $p_T$ distribution of the
leading jet is that the corrections are slightly larger. The results
are shown in the appendix (\Fig{fig:3jpT2} and \Fig{fig:3jpT3}).
\begin{figure}[t]
  \begin{center}
    \includegraphics[width=\columnwidth]{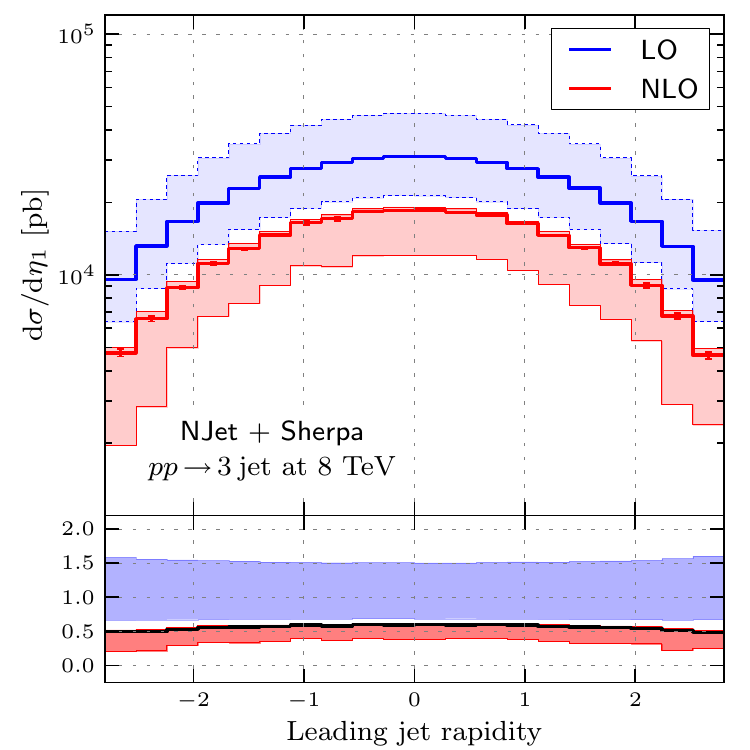}
  \end{center}
  \caption{Rapidity distribution in $pp\to3$ jets for the leading jet at the LHC with a centre-of-mass energy of 8 TeV.}
  \label{fig:3jeta1}
\end{figure}
In \Fig{fig:3jeta1} we show the rapidity distribution of the leading jet.
As in the $p_T$ distribution we observe a sizeable correction at NLO
together with an important reduction of the scale uncertainty. Again
the K-factor is rather flat. We have also analysed the rapidity
distribution of the second and third jet. The results look very
similar to \Fig{fig:3jeta1}. The explicit plots are given in the
appendix.

\subsection{Four Jet Production}
\begin{figure}[t]
  \begin{center}
    \includegraphics[width=0.49\textwidth]{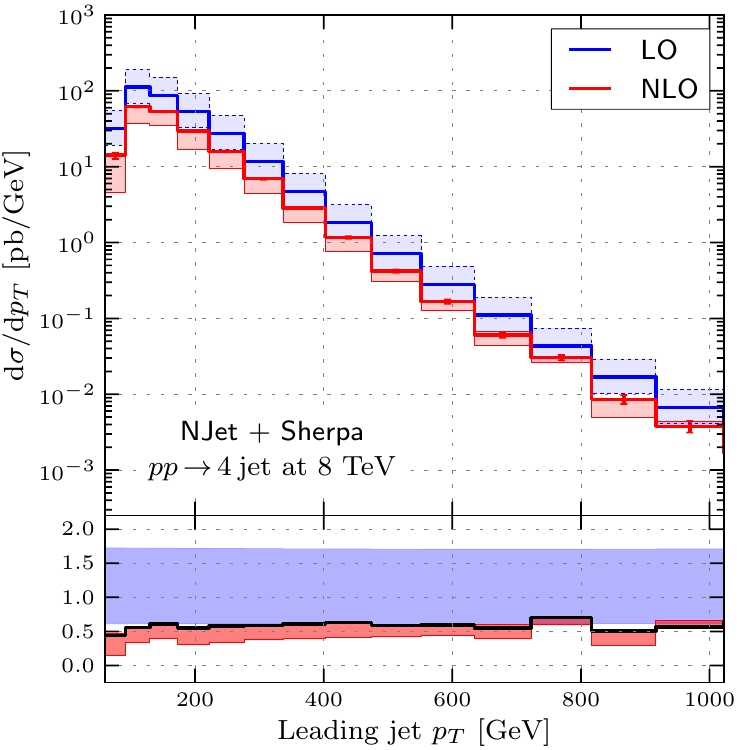}
  \end{center}
  \caption{$p_T$ distribution in $pp\to4$ jets for the leading jet at the LHC with a centre-of-mass energy of 8 TeV.}
  \label{fig:4jpT1}
\end{figure}

\begin{figure}[t]
  \begin{center}
    \includegraphics[width=0.49\textwidth]{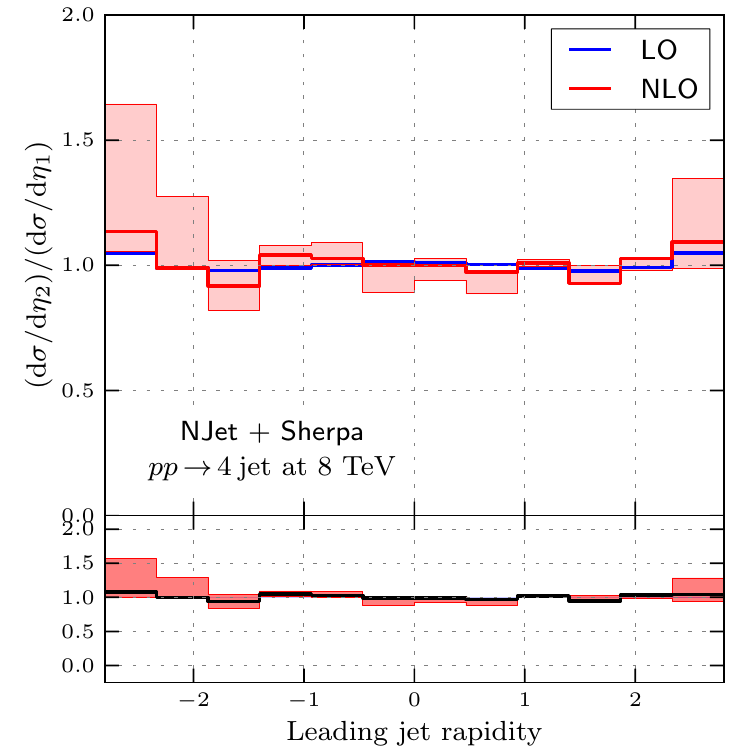}
  \end{center}
  \caption{Ratio $R_2$ of the rapidity distribution for the second jet
    with respect to the leading jet.}
  \label{fig:4jeta12ratio}
\end{figure}

In this subsection we present new results for four jet production at
NLO accuracy in QCD.  Performing a similar analysis as in the three
jet case we find for the four jet cross section at 8 TeV:
\begin{align}
  \sigma_4^{\text{8TeV-LO}} &= 14.36(0.01)^{+10.38}_{-5.6}\: {\rm nb}, \\
  \sigma_4^{\text{8TeV-NLO}} &= 8.15(0.09)^{+0.0}_{-3.24}\: {\rm nb}.
  \label{eq:4jXS}
\end{align}
Similar to the findings for 7 TeV the size of the NLO corrections
amount to a reduction of $-45$\% compared to LO. As far as the scale
dependence is concerned a new feature appears compared to the three
jet rate: The cross sections for $\mu = \hat{H}_T/4$ and $\mu=\hat{H}_T$ do not
bracket the result for $\mu=\hat{H}_T/2$. This is not uncommon for a NLO
calculation. While the LO cross section is decreasing as a function of
the renormalization scale---a direct consequence of the negative value
of the beta function---the NLO scale dependence may develop a plateau
showing a flat scale dependence in a restricted region. This is in
fact what we would consider as an ideal situation. If the central
scale is close to the extrema two additional results obtained by
varying the scale by a factor two up and down will be both larger or
smaller than the central value. This is precisely what happens in case
of the four jet cross section. As central result we quote in
\Eq{eq:4jXS} the result for $\mu=\hat{H}_T/2$. Assuming that this value is
already close to the extrema in the restricted scale range
$\hat{H}_T/4<\mu<\hat{H}_T$ we set the super-script describing the upwards
variation to zero. The sub-script describing the downwards variation
is obtained from $\min(\sigma_4^{\text{8TeV-NLO}}(\mu=\hat{H}_T/4),\sigma_4^{\rm
  8TeV-NLO}(\mu=\hat{H}_T))$.  Using the explicit results for $\mu=\hat{H}_T/4$
and $\mu=\hat{H}_T$
  \begin{align}
    \sigma_4^{\text{8TeV-NLO}}(\mu=\hat{H}_T) &= 7.91(0.05)\: {\rm nb},\\
    \sigma_4^{\text{8TeV-NLO}}(\mu=\hat{H}_T/4) &= 4.91(0.15)\: {\rm nb},
  \end{align}
thus we get the quoted value for the lower scale variation band, $-3.24\: {\rm nb}$.
As in the three jet case moving from
7 TeV collider energy to 8 TeV increases the cross section. For the
four jet cross section the effect is with about 50\% slightly larger
than for the three jet rate. We also mentioned that the relative numerical
uncertainty due to the Monte Carlo integration is of the order of
1\%. The larger uncertainty compared to the three jet cross section
reflects the fact that the integration is much more involved.
Although not a physical observable
it is illustrative to study the contributions from different parton
channels. The result is shown in \Tab{tab:PartonFractions}. Note that
we do not distinguish between quarks and anti-quarks in the initial
state. For example the qq channel includes the contribution from the initial
states $q\qbar', qq', \qbar\qbar'$. Furthermore we do not distinguish
different quark flavours.
\begin{table}[htbp]
  \begin{center}
    \leavevmode
    \begin{tabular}[h]{c|c|c|c|}
      &gg & qg & qq \\ \hline
      relative contribution & $37\%$ & $49\%$ & $14\%$ \\
    \end{tabular}
    \caption{Relative contribution of the different parton channels \mbox{to $pp\to 4$} jets at LO.}
    \label{tab:PartonFractions}
  \end{center}
\end{table}
As can be seen the dominant channels are those with one gluon
and one quark in the initial state. This is a direct consequence of
the sizeable partonic cross section in combination with the largest
parton luminosity. $31\%$ of the $49\%$ from
these channels come from $gu\to u+3g$ and $gd\to d+3g$.  The largest
single process is $gg\to 4g$ which contributes $30\%$.  As in
the three jet case we show in \Fig{fig:4jpT1} the $p_T$ distribution
of the leading jet.
The behaviour follows closely what has been observed in the three jet
case. The scale dependence of the NLO result is significantly reduced
compared to the LO prediction. Again the K-factor is almost constant
over a wide $p_T$ range. We find a negative correction of about 45\%
--- in agreement with the findings for the `inclusive' four jet cross
section. Only for extreme $p_T$ values the K-factor changes
significantly. For small $p_T$ this might again be a consequence of
soft gluon corrections which would require to go beyond fixed order
perturbation theory. At large $p_T$ the uncertainties due to the
numerical integration become important, so that the results become
unreliable. The $p_T$ distributions of the remaining jets follow
pretty much the same pattern. The corresponding K-factor is even more
stable compared to the $p_T$ distribution of the highest $p_T$ jet.
The results are shown in \Fig{fig:4jpT2}, \Fig{fig:4jpT3} and
\Fig{fig:4jpT4} in the appendix. In \Fig{fig:4jeta1}
the rapidity distribution of the leading jet is shown. The qualitative
findings are similar to the $p_T$ distribution: Reduction of the NLO
cross sections by about 45--50\%, significant reduction of the scale
dependence and a K-factor which is almost constant over the phase
space sampled by the distributions. Looking into the rapidity distributions
of the remaining jets we find the same picture.
\begin{figure}[t]
  \begin{center}
    \includegraphics[width=0.49\textwidth]{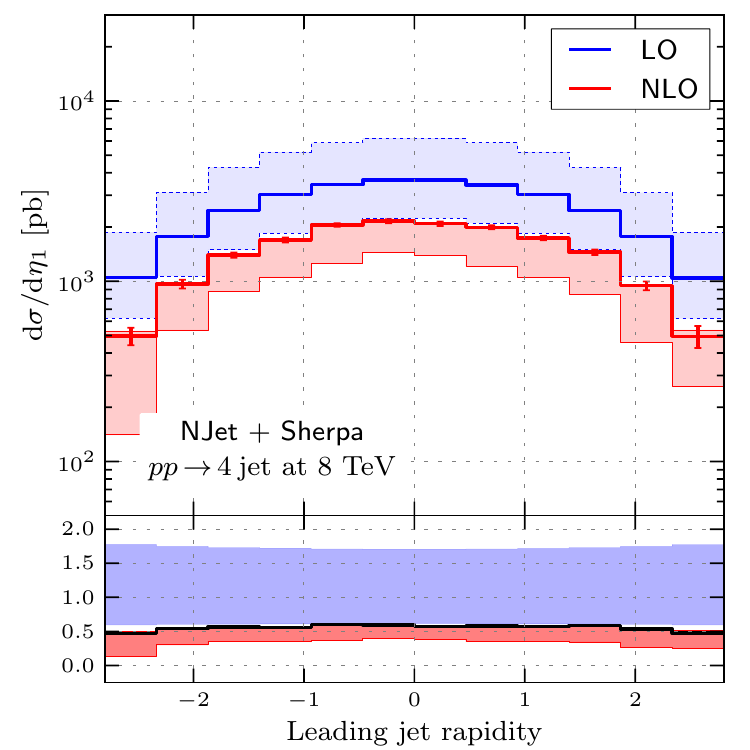}
  \end{center}
  \caption{Rapidity distribution in $pp\to4$ jets for the leading jet
    at the LHC with a centre-of-mass energy of 8 TeV.}
  \label{fig:4jeta1}
\end{figure}
Again we refer to the appendix for the corresponding plots.  We notice
that the rapidity distributions of the different jets look remarkable
similar. To study this in more detail we investigated the ratio
\begin{equation}
  R_{j}=\frac{d\sigma_4}{d\eta_j}\bigg/\frac{d\sigma_4}{d\eta_1}.
\end{equation}
The result for $R_2$ is shown in
\Fig{fig:4jeta12ratio}.
Two important aspects are visible: First of all the ratio is
remarkably stable with respect to perturbative corrections.  Within
the numerical uncertainties LO and NLO results are in perfect
agreement. We also show the effects due to scale variation. However
since in the ratio the leading power in \as\ cancels the scale
variation is not necessarily a reliable estimate of the theoretical
uncertainty. An alternative way to assess the effect of higher order
corrections would be to compare the ratio expanded in \as\ with the
naive ratio where we just divide the predictions for the numerator and
the denominator. However since we use a dynamical scale setting
procedure this is not possible. The second important observation is
that to good approximation we have $R_2\approx 1$ consistent with the
naive expectation. The results for $R_3$ and $R_4$ look very similar.
Experimentally a measurement of the different ratios could be used
to validate detector efficiencies and to further constrain the jet
energy scale.

\begin{figure}[t]
  \begin{center}
    \includegraphics[width=\columnwidth]{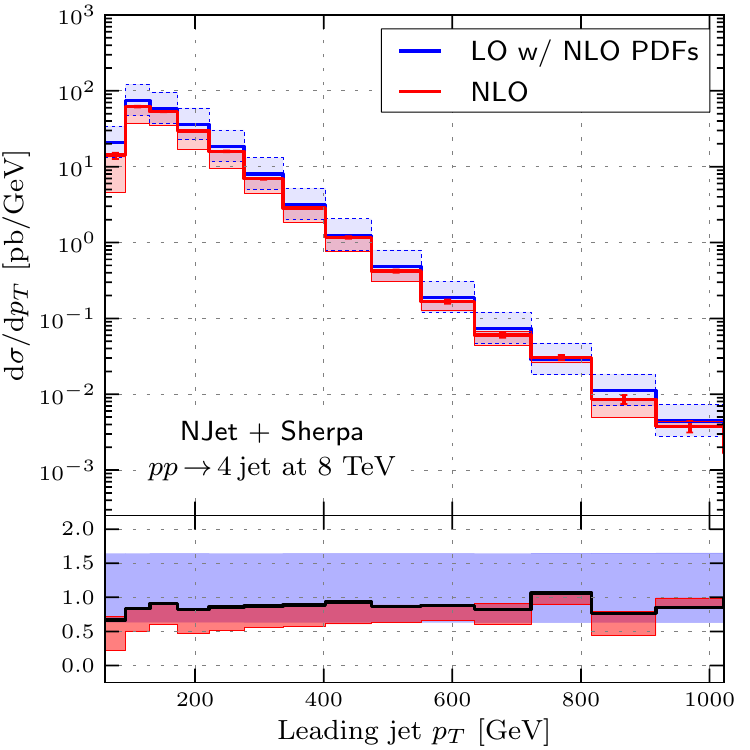}
  \end{center}
 \caption{$p_T$ distribution of the leading jet
   in $pp\to4$ jets using NLO PDFs in the evaluation of the LO cross
   section.}
  \label{fig:4jpT1N}
\end{figure}

\begin{figure}[t]
  \begin{center}
    \includegraphics[width=\columnwidth]{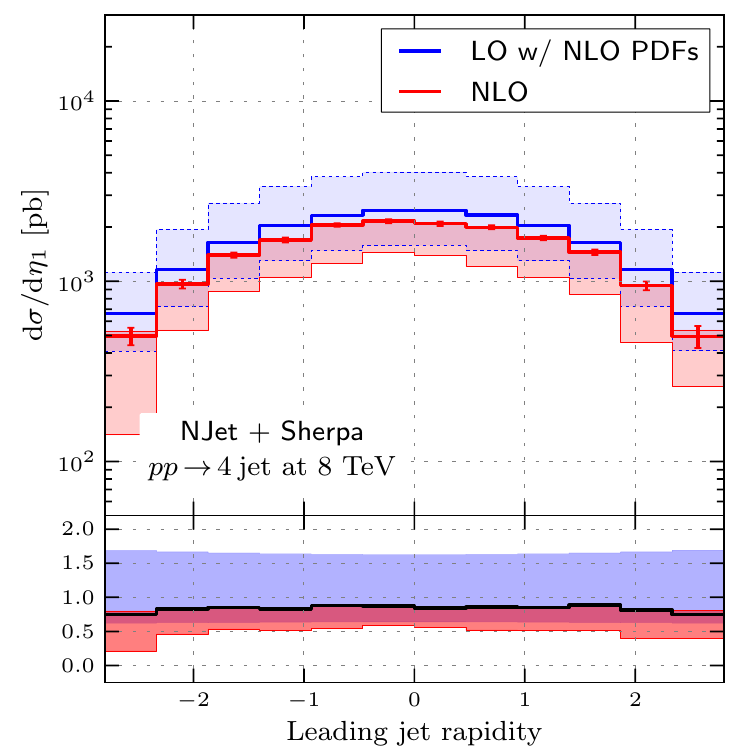}
  \end{center}
  \caption{Normalized eta distribution of the leading jet
    in $pp\to4$ jets using NLO PDFs in the evaluation of the LO cross
    section. }
  \label{fig:4jeta1N}
\end{figure}

The stability of the results shown in \Fig{fig:4jeta12ratio} is
related to the fact that in the ratio the almost constant K-factor
cancels out. In general it might be beneficial to study normalized
distributions since in the ratio many uncertainties may cancel. This
is evidently true for \as\ but should also hold to some extent for
uncertainties due to the parton distribution functions.
\begin{figure}[t]
  \begin{center}
    \includegraphics[width=\columnwidth]{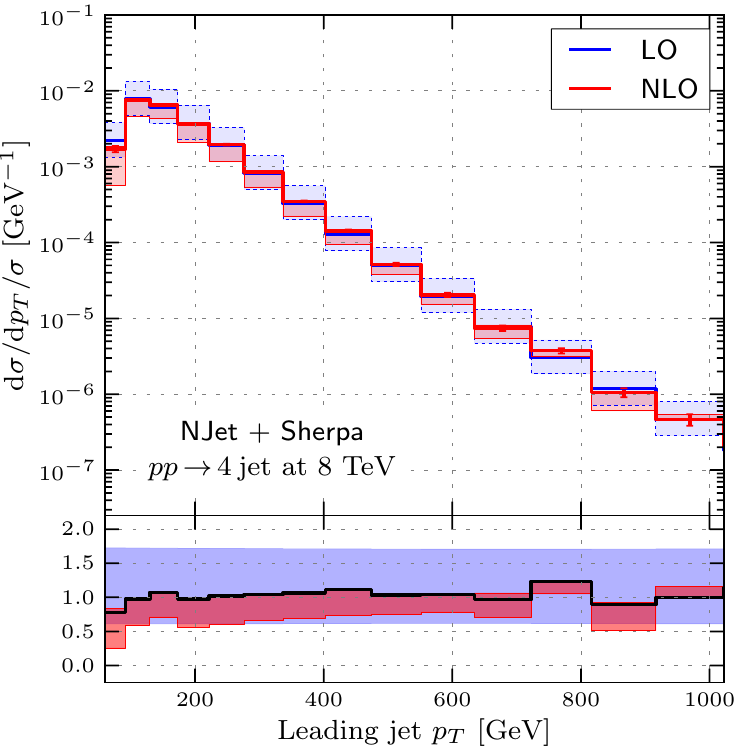}
  \end{center}
 \caption{Normalized $p_T$ distribution of the leading jet
   in $pp\to4$ jets.}
  \label{fig:4jpT1R}
\end{figure}
  \begin{figure}[t]
  \begin{center}
    \includegraphics[width=\columnwidth]{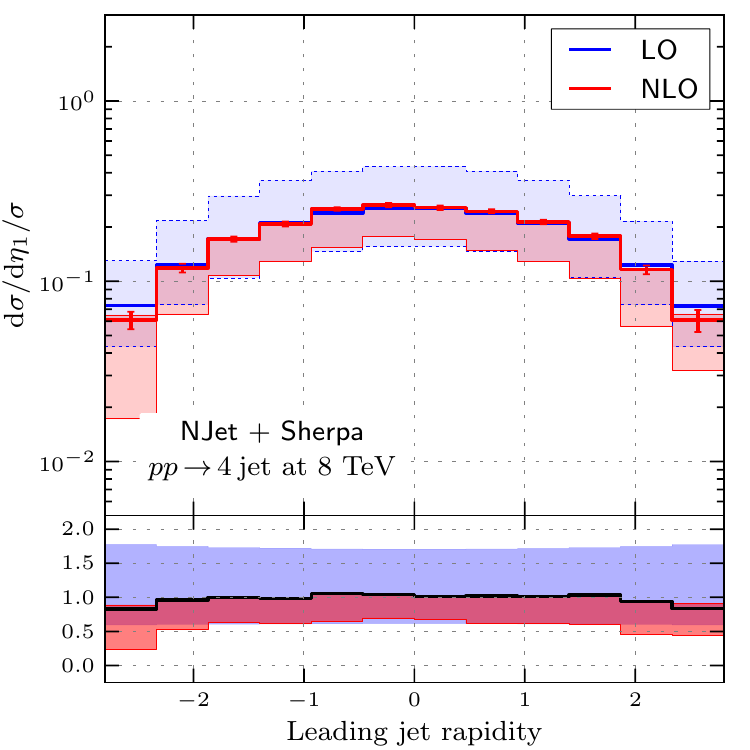}
  \end{center}
  \caption{Normalized eta distribution of the leading jet
    in $pp\to4$ jets.}
  \label{fig:4jeta1R}
\end{figure}
In \Fig{fig:4jpT1R} and \Fig{fig:4jeta1R} we show the normalized
distributions for the leading jet. 
Compared to the un-normalized distributions the size of the corrections
is reduced. In the rapidity distribution for example the K-factor becomes
close to one. Again we stress that the scale uncertainty does not
necessarily provide a reliable estimate of theoretical uncertainty.
We mentioned in the description
of the numerical setup that the LO PDFs come with a rather large value
for \as. Observing the sizeable NLO corrections --- about 40\% in case of
the three jet cross sections and 45\% for the four jet cross
section --- one may ask how much of the corrections are actually due to the shift
in $\as$. Furthermore it would be interesting to disentangle the
perturbative corrections of the hard scattering from the change of the
PDFs when moving from LO to NLO. To do so we show in \Fig{fig:4jpT1N}
and \Fig{fig:4jeta1N} results where we have used NLO PDFs together
with the respective \as\ value to evaluate the LO cross sections.
We see that in this way the size of the NLO corrections is roughly
divided by two. We also observe that LO and NLO predictions
overlap taking the scale uncertainty as an uncertainty estimate of
uncalculated higher orders.
From a phenomenological point of view we are lead to
the conclusion that using the NLO setup in the evaluation of the LO
cross sections gives a much better approximation to what happens at
NLO accuracy compared to the default setup with LO PDFs. This is a
valuable information for the experimental analysis in cases where the
NLO corrections are not available or take to long to be evaluated.  In
general it would be interesting to investigate whether the
observations we made here in the case of the four jet cross section
hold true for a wider class of NLO processes. It is conceivable that a
similar procedure also works for other processes. The reasoning behind
the large LO \as\ value is to account for the NLO corrections to the
matrix elements which are not considered in the LO partonic cross
sections. However for this to work for a large variety of different
processes would require a universal K-factor for NLO corrections.
Since K-factors can be rather different for different processes there
is an obvious limitation of the procedure. Naively one could expect
that this becomes in particular true when cross sections are
considered which are of different order in \as\ compared to the cross
sections which enter PDF fits. In such cases using an NLO setting in
the evaluation of LO cross sections may provide a better approximation
to the `full' (NLO) answer.

\subsection{Comparison of three and four jet production}
As reference value we have calculated also the two jet cross section
using the same setup as before:
 \begin{align}
  \sigma_2^{\text{8TeV-LO}} &= 1234.9(1.2)\: {\rm nb}, \\
  \sigma_2^{\text{8TeV-NLO}} &= 1524.9(2.8)\:  {\rm nb}.
  \label{eq:2jXS}
\end{align}
Combining (not quite consistent in \as\ and ignoring soft gluon
resummation) the results of \Eq{eq:3jXS}, \Eq{eq:4jXS}, and
\Eq{eq:2jXS} we estimate the total jet cross section to be of the
order of 1600~nb. We thus obtain for the 2~:~3~:~4 jet ratios:
1~:~0.05~:~0.005. Only 5\% of the multi-jet events are three jet
events. The four jet topology is further reduced by a factor 1/10.  It
is interesting to study how the ratio between four jet production and
three jet production behaves as function of the leading jet $p_T$.
Similar to what has been shown in \Ref{Bern:2011ep} we analyse
the ratio
\begin{equation}
  \frac{d\sigma_4}{dp_T }\bigg/ \frac{d\sigma_3}{dp_T }.
\end{equation}
\begin{figure}[t]
  \begin{center}
    \includegraphics[width=\columnwidth]{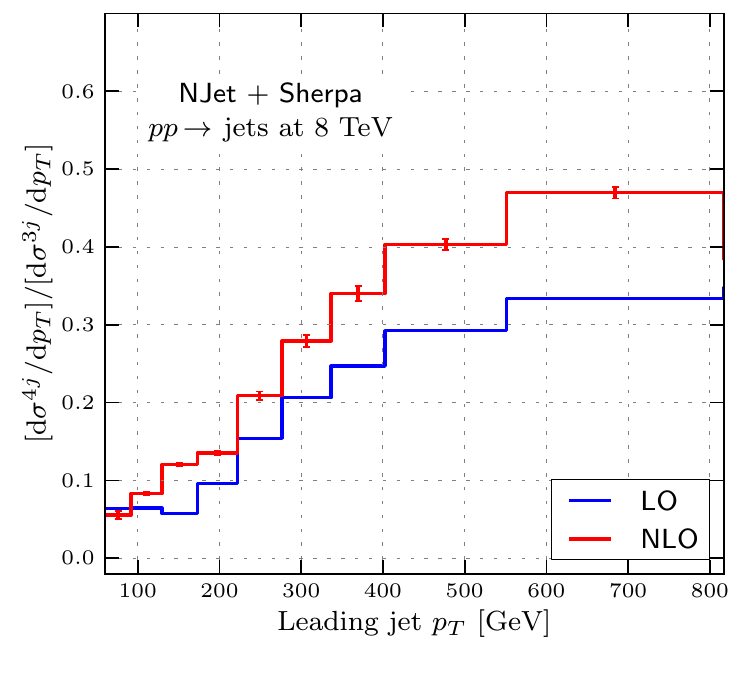}
  \end{center}
  \caption{Comparison of three and four jet production. }
  \label{fig:JetRatios}
\end{figure}
We observe in \Fig{fig:JetRatios} that the reduction of the four jet
cross section with respect to the three jet cross section is mostly
due to the low $p_T$ region. At large $p_T$ the fraction of four jet
events with respect to three jet events raises. In leading order the
fraction is about 1/3 at 800~GeV while in NLO it is close to 1/2. With the
exception of the low $p_T$ region we find that the K-factor is rather
constant and takes a value of about 1.4.
Note that we have not shown the scale variation in
\Fig{fig:JetRatios}. Since in the ratio the scale dependence would
largely cancel the scale dependence will not provide a reliable
estimate of the uncalculated higher orders. Note that the results
shown here slightly differ from what has been shown in
\Ref{Bern:2011ep}. We expect the differences to be a consequence of
the different bin sizes and the different $R$ value used in the
anti-kt jet algorithm.

\section{Conclusions}
We have presented a study of three and four jet production at the LHC
running at a centre-of-mass energy of 8 TeV.  The virtual corrections
were efficiently evaluated using an on-shell unitarity based method
implemented in the publicly available \NJet C++ library. We have cross
checked the virtual corrections for individual phase space points as
far as possible with existing results. We find complete agreement. The
complicated structure of real radiation processes was treated within
the Catani-Seymour dipole subtraction method. We used the Sherpa Monte
Carlo program for this. As an important cross check of our approach we
reproduced results for LHC at 7 TeV published recently by the \BlackHat
collaboration. We find perfect agreement. We stress that this
comparison represents the first independent check of the results given
in \Ref{Bern:2011ep}.
The calculation documented in this article nicely illustrates
the performance of the \NJet library to evaluate QCD one-loop
corrections.

For the three and four jet cross sections we find the
NLO corrections behave similarly to those obtained at 7 TeV. The NLO
corrections reduce the LO cross sections by about 40--50\% depending
on the jet multiplicity. Compared to the 7 TeV results the cross sections
are increased by about 50\% due to the larger collider energy and the
related increase in the parton fluxes.
We have also studied differential distributions. The dynamical scale
setting procedure results in an almost constant K-factor leading to
very stable results for the normalized distributions. To pin down the
origin of the large negative corrections observed in the cross
sections and in the unnormalized distributions we have analysed the impact
of the LO PDFs used in the LO predictions. Using an NLO setup in the
LO calculation shows that the large negative corrections are largely
due to the change in the PDFs when going from LO to NLO with the \as\
value being of particular importance.

The corresponding data for all the plots shown in this article is
tabulated in appendix \ref{sec:tables}. 

We stress that the results presented here are in fixed order
perturbation theory. Large corrections at low $p_T$ due to soft gluon
emission may require to go beyond fixed order in perturbation theory
to achieve a good agreement with the data.  Significant progress has
been obtained in the matching of NLO calculations with parton shower
predictions.  For a number of phenomenologically important processes
matched results are now available and the automation is pushed forward
in two different frameworks ((a)MC@NLO
\cite{Frixione:2002ik,Frixione:2007vw,Frederix:2011zi} and POWHEG
\cite{Nason:2004rx,Alioli:2010xd}). Studies of di-jet production at NLO
with parton showering have recently been performed using both methods
\cite{Alioli:2010xa,Hoeche:2012fm}. Since the \NJet library we
developed to perform the calculation presented here is publicly
available all required ingredients to include four jet production in
(a)MC@NLO or POWHEG are available. We also note that the \NJet library
includes the virtual corrections required for the  calculation of the
five jet cross section at NLO accuracy.

\section*{Acknowledgments}

We are very grateful to Fabio Maltoni, Rikkert Frederix and Marco Zaro
for useful discussions and testing the \NJet code within the aMC@NLO
framework. We would also would like to thank Frank Krauss and Stefan
Hoeche for assistance with {\sc Sherpa}. Special thanks go to Alberto
Guffanti for numerous helpful comments. This work is supported by the
Helmholtz Gemeinschaft under contract HA-101 (Alliance Physics at the
Terascale) and by the European Commission through contract
PITN-GA-2010-264564 (LHCPhenoNet). We would also like to thank the DESY,
Zeuthen theory group for providing computer resources.

\appendix
\section{Additional differential distributions for three and four 
jet production}
In this section we show results for the transverse momentum  and
rapidity distributions of the subleading jets. Since they show a
behaviour rather similar to the leading jet they were not discussed in
detail in the main text. 
\begin{figure}[t]
  \begin{center}
    \includegraphics[width=\columnwidth]{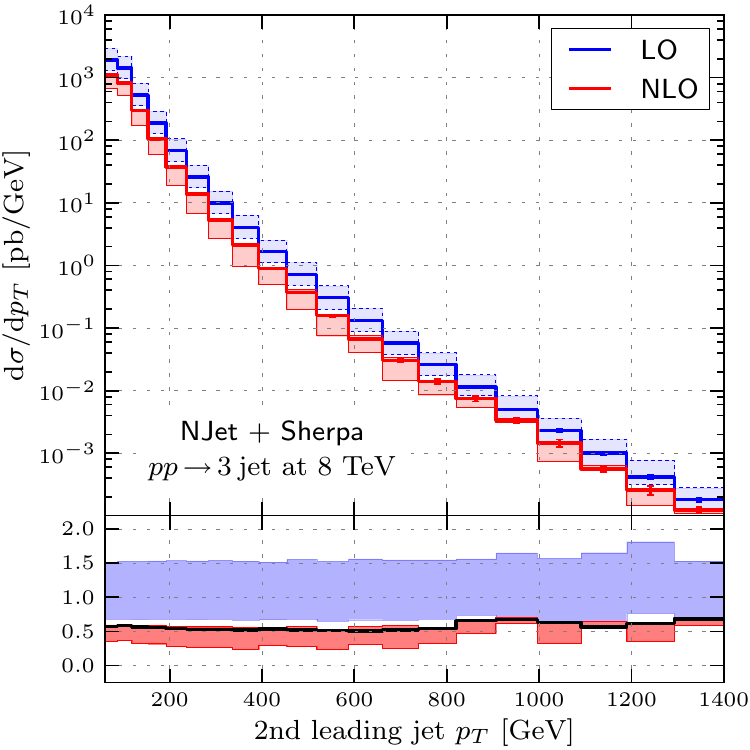}
  \end{center}
  \caption{$p_T$ distribution in $pp\to3$ jets for the $2^{\rm nd}$
    jet at the LHC with a centre-of-mass energy of 8 TeV.}
  \label{fig:3jpT2}
\end{figure}

\begin{figure}[t]
  \begin{center}
    \includegraphics[width=\columnwidth]{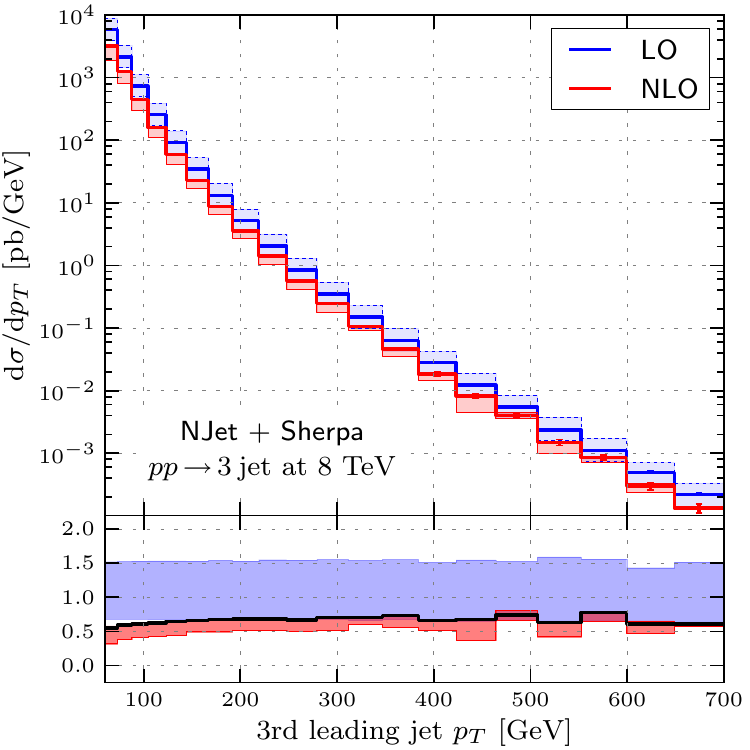}
  \end{center}
  \caption{$p_T$ distribution in $pp\to3$ jets for the $3^{\rm rd}$
    jet at the LHC with a centre-of-mass energy of 8 TeV.}
  \label{fig:3jpT3}
\end{figure}

\begin{figure}[t]
  \begin{center}
    \includegraphics[width=\columnwidth]{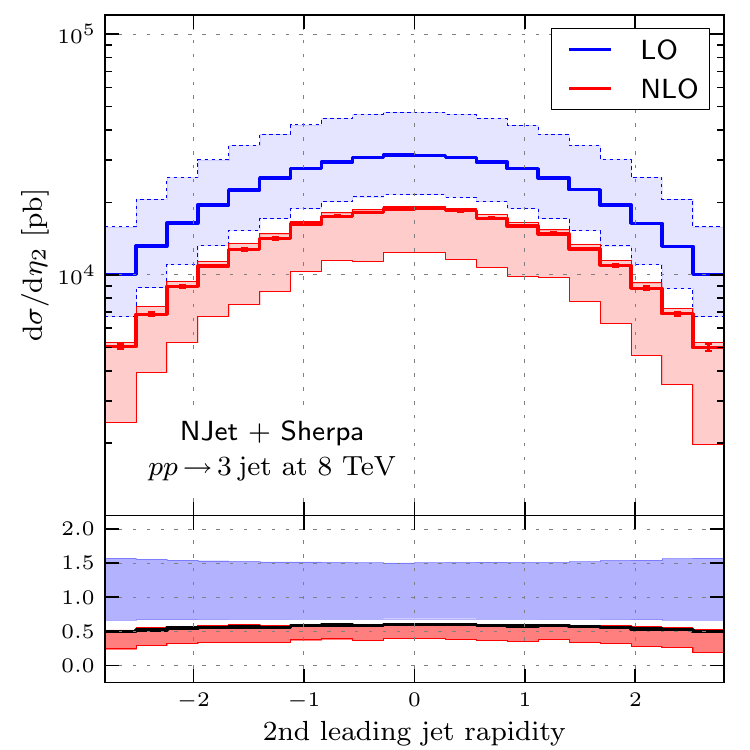}
  \end{center}
  \caption{Rapidity distribution in $pp\to3$ jets for the $2^{\rm nd}$ jet at the LHC with a centre-of-mass energy of 8 TeV.}
  \label{fig:3jeta2}
\end{figure}

\begin{figure}[t]
  \begin{center}
    \includegraphics[width=\columnwidth]{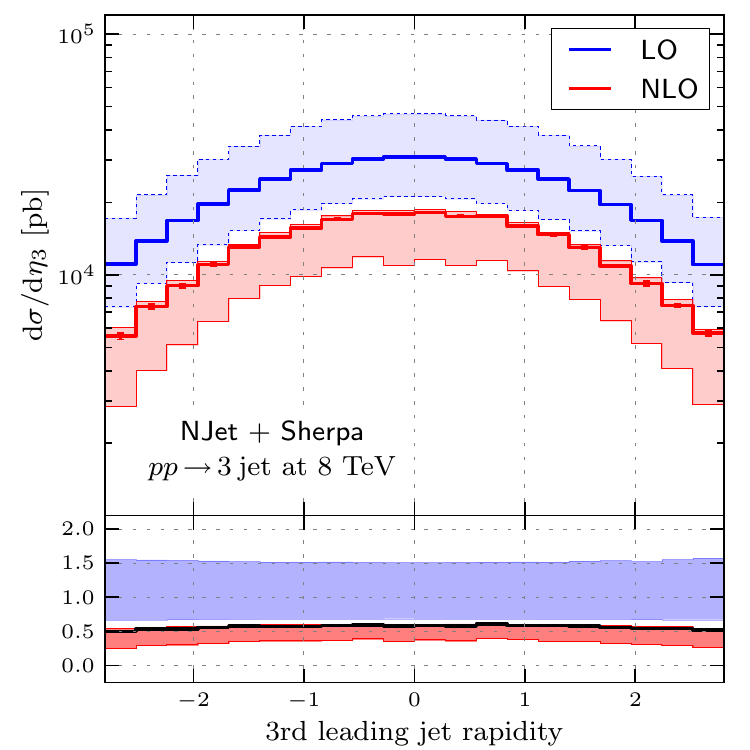}
  \end{center}
  \caption{Rapidity distribution in $pp\to3$ jets for the $3^{\rm rd}$ jet at the LHC with a centre-of-mass energy of 8 TeV.}
  \label{fig:3jeta3}
\end{figure}

\begin{figure}[t]
  \begin{center}
    \includegraphics[width=0.49\textwidth]{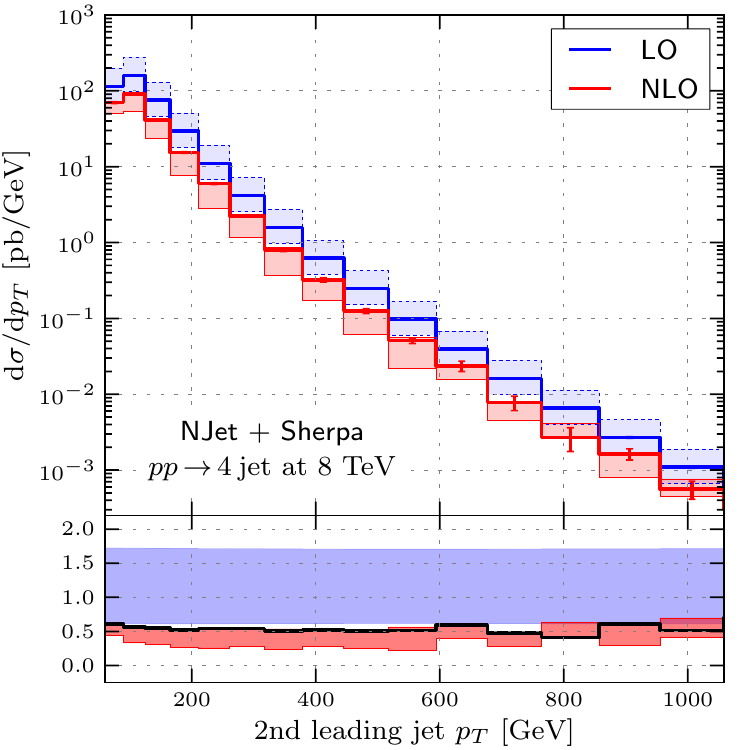}
  \end{center}
  \caption{$p_T$ distribution in $pp\to4$ jets for the $2^{\rm nd}$ jet at the LHC with a centre-of-mass energy of 8 TeV.}
  \label{fig:4jpT2}
\end{figure}

\begin{figure}[t]
  \begin{center}
    \includegraphics[width=0.49\textwidth]{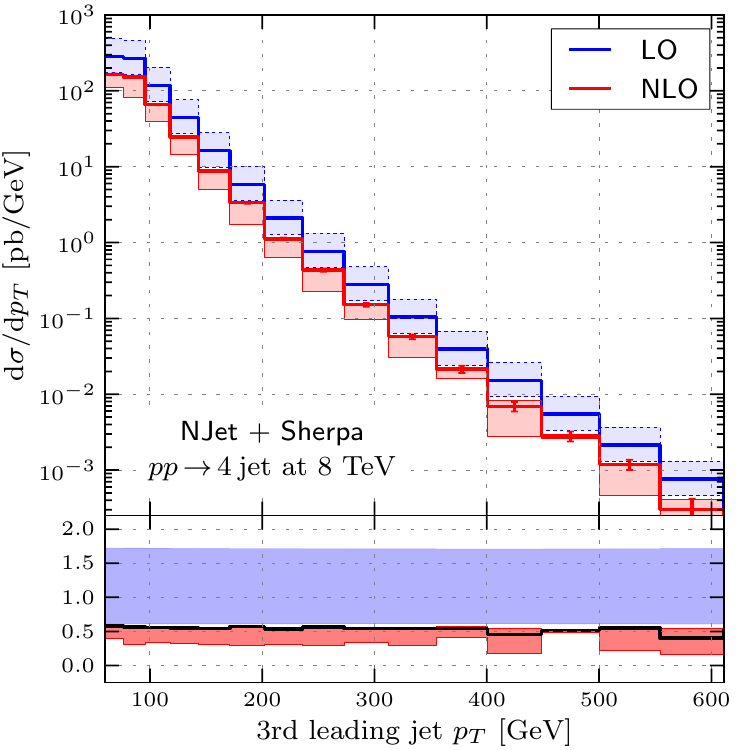}
  \end{center}
  \caption{$p_T$ distribution in $pp\to4$ jets for the $3^{\rm rd}$ jet at the LHC with a centre-of-mass energy of 8 TeV.}
  \label{fig:4jpT3}
\end{figure}

\begin{figure}[t]
  \begin{center}
    \includegraphics[width=0.49\textwidth]{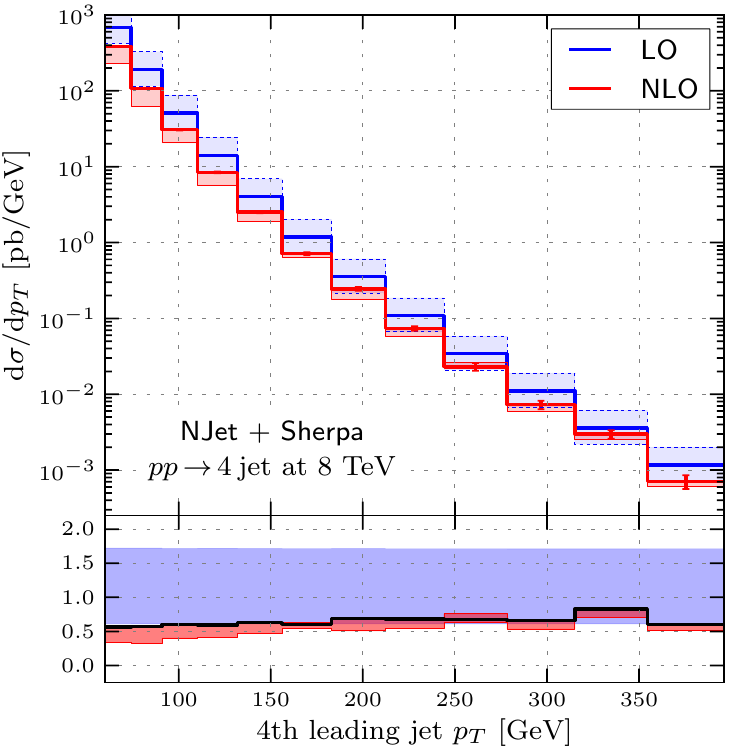}
  \end{center}
  \caption{$p_T$ distribution in $pp\to4$ jets for the $4^{\rm th}$ jet at the LHC with a centre-of-mass energy of 8 TeV.}
  \label{fig:4jpT4}
\end{figure}

\begin{figure}[t]
  \begin{center}
    \includegraphics[width=0.49\textwidth]{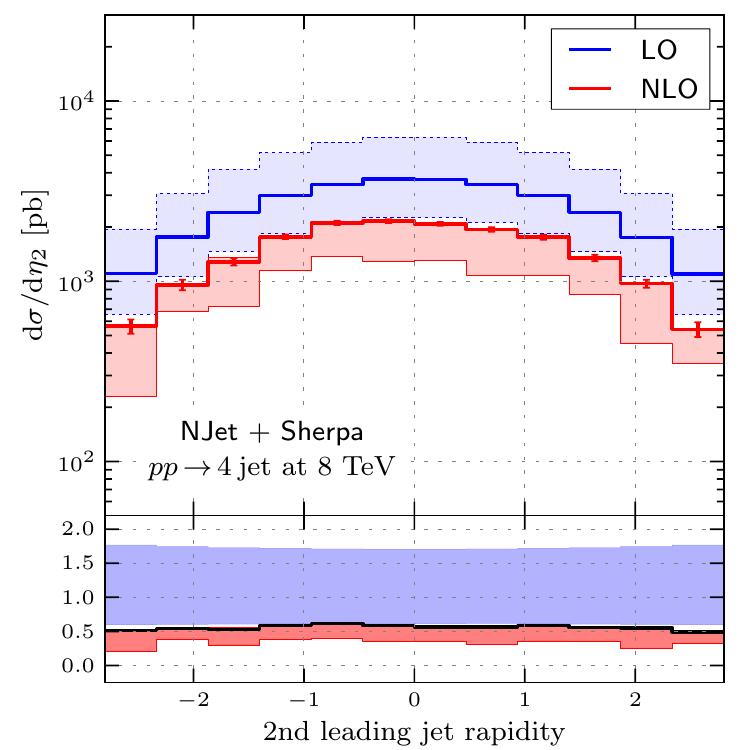}
  \end{center}
  \caption{Rapidity distribution in $pp\to4$ jets for the $2^{\rm nd}$ jet at the LHC with a centre-of-mass energy of 8 TeV.}
  \label{fig:4jeta2}
\end{figure}

\begin{figure}[t]
  \begin{center}
    \includegraphics[width=0.49\textwidth]{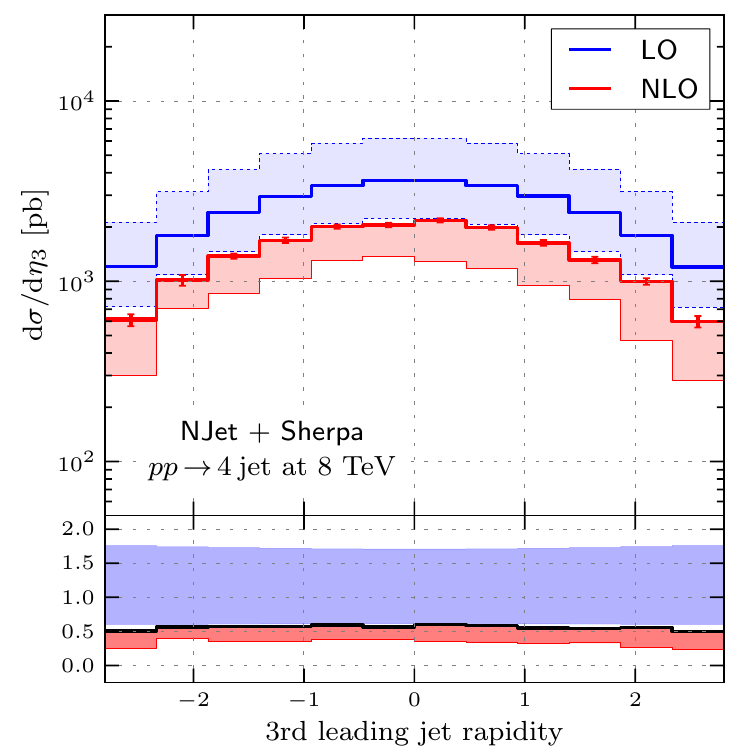}
  \end{center}
  \caption{Rapidity distribution in $pp\to4$ jets for the $3^{\rm rd}$ jet at the LHC with a centre-of-mass energy of 8 TeV.}
  \label{fig:4jeta3}
\end{figure}

\begin{figure}[t]
  \begin{center}
    \includegraphics[width=0.49\textwidth]{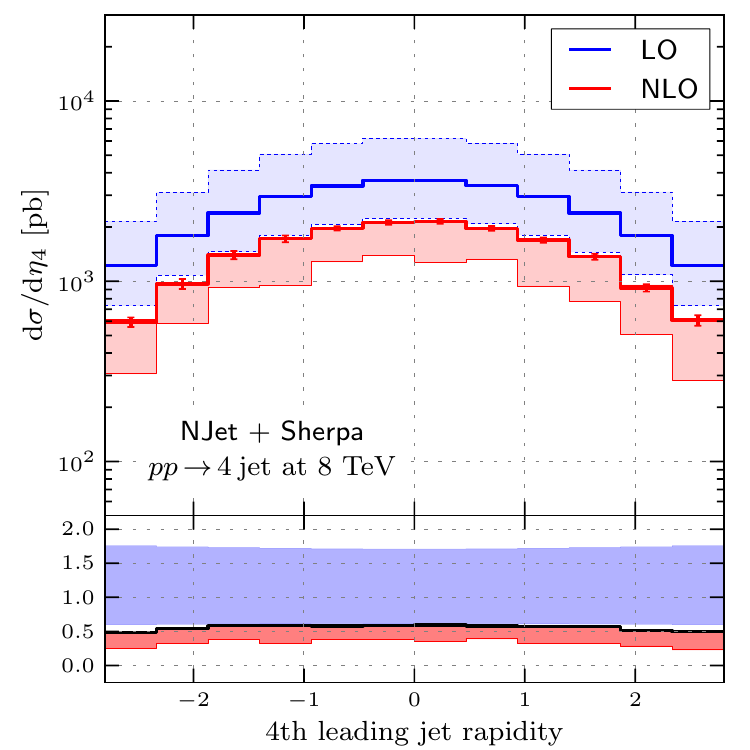}
  \end{center}
  \caption{Rapidity distribution in $pp\to4$ jets for the $4^{\rm th}$ jet at the LHC with a centre-of-mass energy of 8 TeV.}
  \label{fig:4jeta4}
\end{figure}
\clearpage

\section{Tables of differential distributions for three jet
  production}
\label{sec:tables}
For future experimental and theoretical studies it might be useful to
present the data shown in the figures also in numerical form. In the
following we give for all the plots shown in the article the results
obtained in the numerical integration.  
\input{hist_tables.tex}

\end{document}

%% file: hist_tables.tex

\begin{table}\centering
  \begin{tabular}{ccc}
    \hline \\
    $p_{T,1}$ (GeV) & LO (pb) & NLO (pb) \\
    \hline
   $60.0$ --- $86.8$ & $5606.3$ $(14.6)$ & $4316.1$ $(80.8)$ \\
   $86.8$ --- $117.8$ & $32747.4$ $(28.7)$ & $24813.3$ $(163.1)$ \\
   $117.8$ --- $153.1$ & $43942.3$ $(34.2)$ & $18293.4$ $(219.5)$ \\
   $153.1$ --- $192.6$ & $25597.3$ $(22.2)$ & $13464.8$ $(116.1)$ \\
   $192.6$ --- $236.3$ & $11019.8$ $(13.3)$ & $6705.3$ $(49.8)$ \\
   $236.3$ --- $284.3$ & $4504.2$ $(6.7)$ & $2873.1$ $(21.2)$ \\
   $284.3$ --- $336.5$ & $1854.8$ $(3.6)$ & $1208.5$ $(15.8)$ \\
   $336.5$ --- $392.9$ & $779.9$ $(2.1)$ & $501.9$ $(5.2)$ \\
   $392.9$ --- $453.5$ & $334.4$ $(1.3)$ & $220.6$ $(3.1)$ \\
   $453.5$ --- $518.4$ & $145.8$ $(0.7)$ & $94.6$ $(1.5)$ \\
   $518.4$ --- $587.5$ & $65.8$ $(0.5)$ & $43.2$ $(0.7)$ \\
   $587.5$ --- $660.9$ & $29.2$ $(0.2)$ & $19.3$ $(0.4)$ \\
   $660.9$ --- $738.5$ & $13.5$ $(0.2)$ & $9.1$ $(0.3)$ \\
   $738.5$ --- $820.3$ & $6.10$ $(0.09)$ & $3.9$ $(0.1)$ \\
   $820.3$ --- $906.3$ & $2.84$ $(0.06)$ & $2.0$ $(0.1)$ \\
   $906.3$ --- $996.6$ & $1.29$ $(0.03)$ & $0.89$ $(0.06)$ \\
   $996.6$ --- $1091.1$ & $0.60$ $(0.02)$ & $0.49$ $(0.03)$ \\
   $1091.1$ --- $1189.8$ & $0.267$ $(0.009)$ & $0.20$ $(0.01)$ \\
   $1189.8$ --- $1292.8$ & $0.123$ $(0.005)$ & $0.09$ $(0.02)$ \\
   $1292.8$ --- $1400.0$ & $0.054$ $(0.003)$ & $0.043$ $(0.004)$ \\
\hline
  \end{tabular}
\caption{Table for the $p_T$ distribution of the 1$^\text{st}$ jet in $pp\to 3$ jets.}
\end{table}

\begin{table}\centering
  \begin{tabular}{ccc}
    \hline \\
    $p_{T,2}$ (GeV) & LO (pb) & NLO (pb) \\
    \hline
   $60.0$ --- $86.8$ & $51490.8$ $(39.9)$ & $29576.7$ $(204.4)$ \\
   $86.8$ --- $117.8$ & $43903.6$ $(30.1)$ & $25597.6$ $(174.4)$ \\
   $117.8$ --- $153.1$ & $18638.5$ $(17.7)$ & $10495.8$ $(97.3)$ \\
   $153.1$ --- $192.6$ & $7441.0$ $(9.4)$ & $4129.0$ $(38.7)$ \\
   $192.6$ --- $236.3$ & $2995.0$ $(5.0)$ & $1617.8$ $(19.5)$ \\
   $236.3$ --- $284.3$ & $1243.1$ $(3.0)$ & $660.4$ $(10.7)$ \\
   $284.3$ --- $336.5$ & $522.6$ $(1.7)$ & $276.0$ $(3.6)$ \\
   $336.5$ --- $392.9$ & $229.9$ $(1.2)$ & $118.9$ $(2.2)$ \\
   $392.9$ --- $453.5$ & $101.6$ $(0.7)$ & $54.2$ $(1.2)$ \\
   $453.5$ --- $518.4$ & $46.3$ $(0.4)$ & $24.1$ $(0.7)$ \\
   $518.4$ --- $587.5$ & $21.4$ $(0.2)$ & $10.9$ $(0.4)$ \\
   $587.5$ --- $660.9$ & $9.7$ $(0.2)$ & $4.9$ $(0.3)$ \\
   $660.9$ --- $738.5$ & $4.46$ $(0.09)$ & $2.3$ $(0.1)$ \\
   $738.5$ --- $820.3$ & $2.13$ $(0.06)$ & $1.2$ $(0.1)$ \\
   $820.3$ --- $906.3$ & $0.98$ $(0.04)$ & $0.65$ $(0.07)$ \\
   $906.3$ --- $996.6$ & $0.45$ $(0.02)$ & $0.30$ $(0.03)$ \\
   $996.6$ --- $1091.1$ & $0.22$ $(0.01)$ & $0.14$ $(0.02)$ \\
   $1091.1$ --- $1189.8$ & $0.099$ $(0.006)$ & $0.056$ $(0.007)$ \\
   $1189.8$ --- $1292.8$ & $0.043$ $(0.003)$ & $0.027$ $(0.004)$ \\
   $1292.8$ --- $1400.0$ & $0.020$ $(0.001)$ & $0.013$ $(0.002)$ \\
\hline
  \end{tabular}
\caption{Table for the $p_T$ distribution of the 2$^\text{nd}$ jet in $pp\to 3$ jets.}
\end{table}

\begin{table}\centering
  \begin{tabular}{ccc}
    \hline \\
    $p_{T,3}$ (GeV) & LO (pb) & NLO (pb) \\
    \hline
   $60.0$ --- $72.8$ & $74805.4$ $(46.5)$ & $41272.9$ $(208.2)$ \\
   $72.8$ --- $87.6$ & $31467.0$ $(23.7)$ & $18659.0$ $(90.8)$ \\
   $87.6$ --- $104.5$ & $12270.2$ $(12.5)$ & $7475.7$ $(50.6)$ \\
   $104.5$ --- $123.3$ & $4819.1$ $(6.4)$ & $3008.2$ $(23.0)$ \\
   $123.3$ --- $144.2$ & $1928.3$ $(3.6)$ & $1244.0$ $(14.5)$ \\
   $144.2$ --- $167.1$ & $789.0$ $(1.9)$ & $523.2$ $(6.7)$ \\
   $167.1$ --- $192.0$ & $326.9$ $(1.1)$ & $219.4$ $(2.7)$ \\
   $192.0$ --- $219.0$ & $139.4$ $(0.6)$ & $95.4$ $(2.8)$ \\
   $219.0$ --- $248.0$ & $59.4$ $(0.3)$ & $40.6$ $(0.8)$ \\
   $248.0$ --- $278.9$ & $26.0$ $(0.2)$ & $17.4$ $(0.7)$ \\
   $278.9$ --- $312.0$ & $11.48$ $(0.09)$ & $8.1$ $(0.2)$ \\
   $312.0$ --- $347.0$ & $5.25$ $(0.06)$ & $3.7$ $(0.1)$ \\
   $347.0$ --- $384.0$ & $2.37$ $(0.03)$ & $1.73$ $(0.09)$ \\
   $384.0$ --- $423.1$ & $1.10$ $(0.02)$ & $0.72$ $(0.04)$ \\
   $423.1$ --- $464.2$ & $0.503$ $(0.010)$ & $0.34$ $(0.03)$ \\
   $464.2$ --- $507.3$ & $0.235$ $(0.005)$ & $0.17$ $(0.01)$ \\
   $507.3$ --- $552.5$ & $0.107$ $(0.003)$ & $0.067$ $(0.007)$ \\
   $552.5$ --- $599.6$ & $0.052$ $(0.002)$ & $0.041$ $(0.004)$ \\
   $599.6$ --- $648.8$ & $0.0244$ $(0.0009)$ & $0.015$ $(0.002)$ \\
   $648.8$ --- $700.0$ & $0.0113$ $(0.0005)$ & $0.007$ $(0.001)$ \\
\hline
  \end{tabular}
\caption{Table for the $p_T$ distribution of the 3$^\text{rd}$ jet in $pp\to 3$ jets.}
\end{table}

\begin{table}\centering
  \begin{tabular}{ccc}
    \hline \\
    $\eta_1$ & LO (pb) & NLO (pb) \\
    \hline
   $-2.8$ --- $(-2.5)$ & $2683.1$ $(16.1)$ & $1328.9$ $(46.8)$ \\
   $-2.5$ --- $(-2.2)$ & $3692.9$ $(15.3)$ & $1842.1$ $(47.8)$ \\
   $-2.2$ --- $(-2.0)$ & $4669.0$ $(12.9)$ & $2477.9$ $(37.6)$ \\
   $-2.0$ --- $(-1.7)$ & $5567.8$ $(12.0)$ & $3122.1$ $(36.3)$ \\
   $-1.7$ --- $(-1.4)$ & $6398.3$ $(11.2)$ & $3597.6$ $(34.4)$ \\
   $-1.4$ --- $(-1.1)$ & $7139.7$ $(10.4)$ & $4093.4$ $(37.6)$ \\
   $-1.1$ --- $(-0.8)$ & $7739.4$ $(10.1)$ & $4613.4$ $(90.7)$ \\
   $-0.8$ --- $(-0.6)$ & $8220.6$ $(10.0)$ & $4790.3$ $(95.9)$ \\
   $-0.6$ --- $(-0.3)$ & $8518.6$ $(9.5)$ & $5139.7$ $(37.9)$ \\
   $-0.3$ --- $(0.0)$ & $8683.8$ $(9.5)$ & $5184.0$ $(40.4)$ \\
   $0.0$ --- $0.3$ & $8696.8$ $(9.6)$ & $5208.2$ $(51.7)$ \\
   $0.3$ --- $0.6$ & $8541.3$ $(9.5)$ & $5073.1$ $(42.3)$ \\
   $0.6$ --- $0.8$ & $8208.1$ $(9.7)$ & $4967.0$ $(37.2)$ \\
   $0.8$ --- $1.1$ & $7730.9$ $(10.1)$ & $4571.0$ $(34.3)$ \\
   $1.1$ --- $1.4$ & $7129.7$ $(10.6)$ & $4090.0$ $(34.6)$ \\
   $1.4$ --- $1.7$ & $6421.9$ $(11.3)$ & $3632.4$ $(44.6)$ \\
   $1.7$ --- $2.0$ & $5580.1$ $(12.1)$ & $3113.5$ $(45.6)$ \\
   $2.0$ --- $2.2$ & $4677.1$ $(13.7)$ & $2536.4$ $(60.3)$ \\
   $2.2$ --- $2.5$ & $3682.1$ $(14.8)$ & $1890.8$ $(59.3)$ \\
   $2.5$ --- $2.8$ & $2670.7$ $(18.5)$ & $1299.0$ $(42.9)$ \\
\hline
  \end{tabular}
\caption{Table for the rapidity distribution of the 1$^\text{st}$ jet in $pp\to 3$ jets.}
\end{table}

\begin{table}\centering
  \begin{tabular}{ccc}
    \hline \\
    $\eta_2$ & LO (pb) & NLO (pb) \\
    \hline
   $-2.8$ --- $(-2.5)$ & $2815.4$ $(13.9)$ & $1413.7$ $(36.2)$ \\
   $-2.5$ --- $(-2.2)$ & $3688.7$ $(12.9)$ & $1921.8$ $(35.9)$ \\
   $-2.2$ --- $(-2.0)$ & $4591.1$ $(11.6)$ & $2511.6$ $(45.0)$ \\
   $-2.0$ --- $(-1.7)$ & $5473.5$ $(11.7)$ & $3047.2$ $(39.3)$ \\
   $-1.7$ --- $(-1.4)$ & $6304.7$ $(11.7)$ & $3578.4$ $(48.0)$ \\
   $-1.4$ --- $(-1.1)$ & $7083.9$ $(11.9)$ & $3967.2$ $(47.7)$ \\
   $-1.1$ --- $(-0.8)$ & $7727.6$ $(11.1)$ & $4556.2$ $(41.9)$ \\
   $-0.8$ --- $(-0.6)$ & $8250.3$ $(11.8)$ & $4901.3$ $(46.6)$ \\
   $-0.6$ --- $(-0.3)$ & $8609.4$ $(11.5)$ & $5078.2$ $(45.0)$ \\
   $-0.3$ --- $(0.0)$ & $8803.3$ $(11.6)$ & $5269.1$ $(59.3)$ \\
   $0.0$ --- $0.3$ & $8788.3$ $(11.3)$ & $5284.6$ $(46.5)$ \\
   $0.3$ --- $0.6$ & $8596.2$ $(11.6)$ & $5176.7$ $(51.4)$ \\
   $0.6$ --- $0.8$ & $8234.9$ $(11.8)$ & $4807.0$ $(48.9)$ \\
   $0.8$ --- $1.1$ & $7732.1$ $(11.5)$ & $4462.6$ $(45.6)$ \\
   $1.1$ --- $1.4$ & $7080.3$ $(12.9)$ & $4145.7$ $(66.0)$ \\
   $1.4$ --- $1.7$ & $6319.7$ $(12.1)$ & $3591.5$ $(43.4)$ \\
   $1.7$ --- $2.0$ & $5468.7$ $(12.0)$ & $3067.9$ $(37.7)$ \\
   $2.0$ --- $2.2$ & $4588.0$ $(12.5)$ & $2462.2$ $(51.4)$ \\
   $2.2$ --- $2.5$ & $3680.5$ $(12.7)$ & $1931.2$ $(40.3)$ \\
   $2.5$ --- $2.8$ & $2815.3$ $(13.7)$ & $1396.6$ $(46.7)$ \\
\hline
  \end{tabular}
\caption{Table for the rapidity distribution of the 2$^\text{nd}$ jet in $pp\to 3$ jets.}
\end{table}

\begin{table}\centering
  \begin{tabular}{ccc}
    \hline \\
    $\eta_3$ & LO (pb) & NLO (pb) \\
    \hline
   $-2.8$ --- $(-2.5)$ & $3102.7$ $(12.3)$ & $1558.4$ $(46.9)$ \\
   $-2.5$ --- $(-2.2)$ & $3877.2$ $(13.3)$ & $2069.5$ $(43.4)$ \\
   $-2.2$ --- $(-2.0)$ & $4704.9$ $(11.2)$ & $2524.2$ $(45.5)$ \\
   $-2.0$ --- $(-1.7)$ & $5507.7$ $(11.5)$ & $3091.2$ $(43.8)$ \\
   $-1.7$ --- $(-1.4)$ & $6294.6$ $(12.4)$ & $3660.3$ $(47.9)$ \\
   $-1.4$ --- $(-1.1)$ & $7005.9$ $(11.4)$ & $4030.4$ $(42.4)$ \\
   $-1.1$ --- $(-0.8)$ & $7628.2$ $(11.7)$ & $4385.8$ $(51.1)$ \\
   $-0.8$ --- $(-0.6)$ & $8117.4$ $(12.0)$ & $4766.5$ $(40.8)$ \\
   $-0.6$ --- $(-0.3)$ & $8464.6$ $(12.2)$ & $5028.8$ $(42.0)$ \\
   $-0.3$ --- $(0.0)$ & $8649.2$ $(12.9)$ & $5012.9$ $(50.5)$ \\
   $0.0$ --- $0.3$ & $8652.7$ $(12.5)$ & $5081.1$ $(47.0)$ \\
   $0.3$ --- $0.6$ & $8467.3$ $(12.3)$ & $4897.5$ $(54.5)$ \\
   $0.6$ --- $0.8$ & $8109.7$ $(12.0)$ & $4927.0$ $(51.2)$ \\
   $0.8$ --- $1.1$ & $7626.6$ $(12.5)$ & $4463.0$ $(51.9)$ \\
   $1.1$ --- $1.4$ & $7005.6$ $(12.5)$ & $4112.4$ $(68.9)$ \\
   $1.4$ --- $1.7$ & $6272.6$ $(11.4)$ & $3639.6$ $(40.9)$ \\
   $1.7$ --- $2.0$ & $5487.1$ $(11.3)$ & $3050.4$ $(43.4)$ \\
   $2.0$ --- $2.2$ & $4717.6$ $(12.3)$ & $2577.2$ $(55.2)$ \\
   $2.2$ --- $2.5$ & $3872.4$ $(12.1)$ & $2090.8$ $(34.1)$ \\
   $2.5$ --- $2.8$ & $3087.8$ $(12.4)$ & $1603.9$ $(37.7)$ \\
\hline
  \end{tabular}
\caption{Table for the rapidity distribution of the 3$^\text{rd}$ jet in $pp\to 3$ jets.}
\end{table}

\clearpage

\section{Tables of differential distributions for four jet production}


\begin{table}\centering
  \begin{tabular}{ccc}
    \hline \\
    $p_{T,1}$ (GeV) & LO (pb) & NLO (pb) \\
    \hline
   $60.0$ --- $92.0$ & $1013.0$ $(3.8)$ & $446.9$ $(41.5)$ \\
   $92.0$ --- $129.6$ & $4204.4$ $(3.3)$ & $2335.6$ $(57.6)$ \\
   $129.6$ --- $172.9$ & $3775.5$ $(2.4)$ & $2303.3$ $(43.1)$ \\
   $172.9$ --- $221.9$ & $2626.8$ $(1.9)$ & $1452.7$ $(31.5)$ \\
   $221.9$ --- $276.5$ & $1498.8$ $(1.4)$ & $867.7$ $(22.6)$ \\
   $276.5$ --- $336.7$ & $707.1$ $(0.8)$ & $417.7$ $(11.1)$ \\
   $336.7$ --- $402.6$ & $308.8$ $(0.4)$ & $187.0$ $(5.1)$ \\
   $402.6$ --- $474.1$ & $131.5$ $(0.3)$ & $83.2$ $(3.1)$ \\
   $474.1$ --- $551.3$ & $55.4$ $(0.1)$ & $32.5$ $(1.6)$ \\
   $551.3$ --- $634.1$ & $23.32$ $(0.08)$ & $13.8$ $(0.9)$ \\
   $634.1$ --- $722.6$ & $9.79$ $(0.05)$ & $5.4$ $(0.5)$ \\
   $722.6$ --- $816.7$ & $4.09$ $(0.03)$ & $2.9$ $(0.2)$ \\
   $816.7$ --- $916.5$ & $1.69$ $(0.02)$ & $0.9$ $(0.1)$ \\
   $916.5$ --- $1021.9$ & $0.705$ $(0.009)$ & $0.40$ $(0.07)$ \\
\hline
  \end{tabular}
\caption{Table for the $p_T$ distribution of the 1$^\text{st}$ jet in $pp\to 4$ jets.}
\end{table}

\begin{table}\centering
  \begin{tabular}{ccc}
    \hline \\
    $p_{T,2}$ (GeV) & LO (pb) & NLO (pb) \\
    \hline
   $60.0$ --- $89.8$ & $3412.4$ $(4.8)$ & $2088.7$ $(58.8)$ \\
   $89.8$ --- $124.8$ & $5607.7$ $(3.1)$ & $3173.1$ $(56.4)$ \\
   $124.8$ --- $165.1$ & $3031.8$ $(1.9)$ & $1670.4$ $(34.1)$ \\
   $165.1$ --- $210.6$ & $1342.6$ $(1.1)$ & $700.9$ $(16.3)$ \\
   $210.6$ --- $261.4$ & $563.4$ $(0.7)$ & $304.7$ $(8.5)$ \\
   $261.4$ --- $317.5$ & $233.4$ $(0.4)$ & $125.9$ $(4.3)$ \\
   $317.5$ --- $378.8$ & $97.5$ $(0.2)$ & $49.1$ $(2.3)$ \\
   $378.8$ --- $445.4$ & $41.5$ $(0.1)$ & $21.5$ $(1.4)$ \\
   $445.4$ --- $517.2$ & $17.77$ $(0.08)$ & $9.0$ $(0.7)$ \\
   $517.2$ --- $594.2$ & $7.59$ $(0.05)$ & $3.9$ $(0.3)$ \\
   $594.2$ --- $676.6$ & $3.27$ $(0.03)$ & $1.9$ $(0.3)$ \\
   $676.6$ --- $764.2$ & $1.42$ $(0.02)$ & $0.7$ $(0.1)$ \\
   $764.2$ --- $857.0$ & $0.61$ $(0.01)$ & $0.25$ $(0.09)$ \\
   $857.0$ --- $955.1$ & $0.264$ $(0.006)$ & $0.16$ $(0.03)$ \\
   $955.1$ --- $1058.4$ & $0.113$ $(0.004)$ & $0.06$ $(0.02)$ \\
\hline
  \end{tabular}
\caption{Table for the $p_T$ distribution of the 2$^\text{nd}$ jet in $pp\to 4$ jets.}
\end{table}

\begin{table}\centering
  \begin{tabular}{ccc}
    \hline \\
    $p_{T,3}$ (GeV) & LO (pb) & NLO (pb) \\
    \hline
   $60.0$ --- $76.4$ & $4670.9$ $(5.1)$ & $2710.4$ $(70.8)$ \\
   $76.4$ --- $95.8$ & $5196.7$ $(3.0)$ & $2941.8$ $(48.8)$ \\
   $95.8$ --- $118.0$ & $2617.1$ $(1.7)$ & $1463.5$ $(25.4)$ \\
   $118.0$ --- $143.2$ & $1123.2$ $(0.9)$ & $619.2$ $(13.2)$ \\
   $143.2$ --- $171.2$ & $455.2$ $(0.5)$ & $247.8$ $(6.7)$ \\
   $171.2$ --- $202.2$ & $180.5$ $(0.3)$ & $103.7$ $(3.0)$ \\
   $202.2$ --- $236.1$ & $71.1$ $(0.1)$ & $37.9$ $(1.6)$ \\
   $236.1$ --- $272.8$ & $28.10$ $(0.08)$ & $15.9$ $(0.8)$ \\
   $272.8$ --- $312.5$ & $11.08$ $(0.04)$ & $6.0$ $(0.4)$ \\
   $312.5$ --- $355.0$ & $4.46$ $(0.02)$ & $2.4$ $(0.2)$ \\
   $355.0$ --- $400.5$ & $1.80$ $(0.01)$ & $1.0$ $(0.1)$ \\
   $400.5$ --- $448.9$ & $0.737$ $(0.008)$ & $0.33$ $(0.05)$ \\
   $448.9$ --- $500.1$ & $0.281$ $(0.004)$ & $0.14$ $(0.02)$ \\
   $500.1$ --- $554.3$ & $0.116$ $(0.002)$ & $0.064$ $(0.010)$ \\
   $554.3$ --- $611.4$ & $0.0434$ $(0.0008)$ & $0.017$ $(0.007)$ \\
\hline
  \end{tabular}
\caption{Table for the $p_T$ distribution of the 3$^\text{rd}$ jet in $pp\to 4$ jets.}
\end{table}

\begin{table}\centering
  \begin{tabular}{ccc}
    \hline \\
    $p_{T,4}$ (GeV) & LO (pb) & NLO (pb) \\
    \hline
   $60.0$ --- $74.2$ & $9745.6$ $(5.8)$ & $5473.4$ $(80.4)$ \\
   $74.2$ --- $91.0$ & $3178.9$ $(2.0)$ & $1811.3$ $(29.9)$ \\
   $91.0$ --- $110.2$ & $984.5$ $(0.8)$ & $591.7$ $(13.6)$ \\
   $110.2$ --- $131.9$ & $307.7$ $(0.4)$ & $182.8$ $(5.2)$ \\
   $131.9$ --- $156.2$ & $97.5$ $(0.2)$ & $61.2$ $(2.0)$ \\
   $156.2$ --- $183.0$ & $31.49$ $(0.07)$ & $19.1$ $(0.9)$ \\
   $183.0$ --- $212.3$ & $10.36$ $(0.03)$ & $7.2$ $(0.4)$ \\
   $212.3$ --- $244.1$ & $3.45$ $(0.02)$ & $2.3$ $(0.2)$ \\
   $244.1$ --- $278.4$ & $1.173$ $(0.008)$ & $0.79$ $(0.09)$ \\
   $278.4$ --- $315.2$ & $0.408$ $(0.004)$ & $0.27$ $(0.03)$ \\
   $315.2$ --- $354.5$ & $0.142$ $(0.002)$ & $0.12$ $(0.02)$ \\
   $354.5$ --- $396.3$ & $0.0491$ $(0.0008)$ & $0.030$ $(0.006)$ \\
\hline
  \end{tabular}
\caption{Table for the $p_T$ distribution of the 4$^\text{th}$ jet in $pp\to 4$ jets.}
\end{table}

\begin{table}\centering
  \begin{tabular}{ccc}
    \hline \\
    $\eta_1$ & LO (pb) & NLO (pb) \\
    \hline
   $-2.8$ --- $(-2.3)$ & $490.6$ $(1.7)$ & $231.7$ $(25.6)$ \\
   $-2.3$ --- $(-1.9)$ & $828.3$ $(1.8)$ & $450.5$ $(24.8)$ \\
   $-1.9$ --- $(-1.4)$ & $1151.0$ $(1.6)$ & $651.9$ $(22.7)$ \\
   $-1.4$ --- $(-0.9)$ & $1413.3$ $(1.8)$ & $790.3$ $(24.8)$ \\
   $-0.9$ --- $(-0.5)$ & $1600.9$ $(1.7)$ & $957.7$ $(21.9)$ \\
   $-0.5$ --- $(0.0)$ & $1698.2$ $(2.2)$ & $1005.6$ $(26.6)$ \\
   $0.0$ --- $0.5$ & $1698.8$ $(2.3)$ & $973.1$ $(27.1)$ \\
   $0.5$ --- $0.9$ & $1600.4$ $(1.6)$ & $929.5$ $(23.6)$ \\
   $0.9$ --- $1.4$ & $1413.9$ $(1.7)$ & $811.2$ $(23.6)$ \\
   $1.4$ --- $1.9$ & $1152.2$ $(1.6)$ & $677.0$ $(24.1)$ \\
   $1.9$ --- $2.3$ & $826.0$ $(1.6)$ & $440.7$ $(25.1)$ \\
   $2.3$ --- $2.8$ & $487.9$ $(1.6)$ & $231.2$ $(32.0)$ \\
\hline
  \end{tabular}
\caption{Table for the rapidity distribution of the 1$^\text{st}$ jet in $pp\to 4$ jets.}
\end{table}

\begin{table}\centering
  \begin{tabular}{ccc}
    \hline \\
    $\eta_2$ & LO (pb) & NLO (pb) \\
    \hline
   $-2.8$ --- $(-2.3)$ & $514.8$ $(1.8)$ & $262.8$ $(23.9)$ \\
   $-2.3$ --- $(-1.9)$ & $819.3$ $(1.4)$ & $445.6$ $(29.8)$ \\
   $-1.9$ --- $(-1.4)$ & $1126.2$ $(1.6)$ & $598.0$ $(27.7)$ \\
   $-1.4$ --- $(-0.9)$ & $1399.4$ $(1.7)$ & $823.5$ $(23.4)$ \\
   $-0.9$ --- $(-0.5)$ & $1602.2$ $(1.6)$ & $984.4$ $(27.8)$ \\
   $-0.5$ --- $(0.0)$ & $1720.2$ $(2.3)$ & $1006.7$ $(23.8)$ \\
   $0.0$ --- $0.5$ & $1716.3$ $(2.5)$ & $972.4$ $(24.2)$ \\
   $0.5$ --- $0.9$ & $1607.2$ $(1.9)$ & $903.8$ $(25.9)$ \\
   $0.9$ --- $1.4$ & $1398.9$ $(1.8)$ & $819.4$ $(26.7)$ \\
   $1.4$ --- $1.9$ & $1126.3$ $(1.6)$ & $628.3$ $(26.2)$ \\
   $1.9$ --- $2.3$ & $818.3$ $(1.5)$ & $452.8$ $(22.6)$ \\
   $2.3$ --- $2.8$ & $512.3$ $(1.4)$ & $252.7$ $(23.4)$ \\
\hline
  \end{tabular}
\caption{Table for the rapidity distribution of the 2$^\text{nd}$ jet in $pp\to 4$ jets.}
\end{table}

\begin{table}\centering
  \begin{tabular}{ccc}
    \hline \\
    $\eta_3$ & LO (pb) & NLO (pb) \\
    \hline
   $-2.8$ --- $(-2.3)$ & $563.2$ $(1.3)$ & $284.2$ $(21.8)$ \\
   $-2.3$ --- $(-1.9)$ & $838.6$ $(1.7)$ & $472.9$ $(32.7)$ \\
   $-1.9$ --- $(-1.4)$ & $1124.0$ $(1.5)$ & $644.1$ $(21.9)$ \\
   $-1.4$ --- $(-0.9)$ & $1381.6$ $(1.7)$ & $787.9$ $(27.2)$ \\
   $-0.9$ --- $(-0.5)$ & $1582.7$ $(2.3)$ & $938.1$ $(24.7)$ \\
   $-0.5$ --- $(0.0)$ & $1693.3$ $(2.7)$ & $957.9$ $(26.6)$ \\
   $0.0$ --- $0.5$ & $1692.3$ $(1.9)$ & $1016.1$ $(27.5)$ \\
   $0.5$ --- $0.9$ & $1580.8$ $(1.8)$ & $928.6$ $(27.0)$ \\
   $0.9$ --- $1.4$ & $1382.8$ $(1.7)$ & $762.1$ $(27.4)$ \\
   $1.4$ --- $1.9$ & $1122.6$ $(1.6)$ & $613.0$ $(25.0)$ \\
   $1.9$ --- $2.3$ & $838.3$ $(1.3)$ & $466.2$ $(19.5)$ \\
   $2.3$ --- $2.8$ & $561.3$ $(1.3)$ & $279.2$ $(20.4)$ \\
\hline
  \end{tabular}
\caption{Table for the rapidity distribution of the 3$^\text{rd}$ jet in $pp\to 4$ jets.}
\end{table}

\begin{table}\centering
  \begin{tabular}{ccc}
    \hline \\
    $\eta_4$ & LO (pb) & NLO (pb) \\
    \hline
   $-2.8$ --- $(-2.3)$ & $572.1$ $(1.5)$ & $277.2$ $(17.0)$ \\
   $-2.3$ --- $(-1.9)$ & $834.6$ $(1.4)$ & $451.9$ $(28.7)$ \\
   $-1.9$ --- $(-1.4)$ & $1118.4$ $(1.5)$ & $652.6$ $(34.0)$ \\
   $-1.4$ --- $(-0.9)$ & $1376.9$ $(1.7)$ & $803.5$ $(34.9)$ \\
   $-0.9$ --- $(-0.5)$ & $1579.3$ $(2.0)$ & $917.5$ $(25.5)$ \\
   $-0.5$ --- $(0.0)$ & $1695.5$ $(2.5)$ & $989.0$ $(27.9)$ \\
   $0.0$ --- $0.5$ & $1695.1$ $(2.0)$ & $1002.0$ $(30.4)$ \\
   $0.5$ --- $0.9$ & $1585.7$ $(2.4)$ & $916.9$ $(28.1)$ \\
   $0.9$ --- $1.4$ & $1379.1$ $(1.8)$ & $789.6$ $(26.3)$ \\
   $1.4$ --- $1.9$ & $1115.4$ $(1.5)$ & $638.2$ $(24.8)$ \\
   $1.9$ --- $2.3$ & $837.7$ $(1.5)$ & $428.9$ $(20.0)$ \\
   $2.3$ --- $2.8$ & $571.7$ $(1.1)$ & $283.1$ $(18.9)$ \\
\hline
  \end{tabular}
\caption{Table for the rapidity distribution of the 4$^\text{th}$ jet in $pp\to 4$ jets.}
\end{table}